\documentstyle[preprint,aps]{revtex}
\begin{document}
\tightenlines
\draft
\title{The Charge and Matter radial distributions of Heavy-Light
mesons calculated on a lattice}

\author{UKQCD Collaboration} 
\author{A.M. Green,
J. Koponen, 
P. Pennanen\thanks{e-mails:anthony.green@helsinki.fi,
jmkopone@rock.helsinki.fi, petrus@hip.fi} }
\address{Department of Physics and Helsinki Institute of Physics\\
P.O. Box 64, FIN--00014 University of Helsinki,Finland}
\author{C. Michael\thanks{e-mail: cmi@liv.ac.uk}}
\address{Department of Mathematical Sciences, University of Liverpool,
 L69 3BX, UK}

\date{\today}
\maketitle
 \begin{abstract}
For a heavy-light meson with a static heavy quark, we can explore the 
light quark  distribution.
The charge and matter radial
distributions of these heavy-light mesons are measured on a
$16^3\times 24$ lattice at $\beta=5.7$ and a hopping parameter
corresponding to a light quark mass about that of the strange quark. 
Both distributions can be well fitted up to 4 lattice  spacings
($r\approx 0.7$ fm) with the exponential form $w_i^2(r)$, where
 $w_i(r)=A\exp(-r/r_i)$.  For the charge(c) and matter(m) distributions
$r_c \approx 0.32(2)$fm and  $r_m \approx 0.24(2)$fm.  We also discuss
the normalisation of the total charge and matter integrated over all
space, finding 1.30(5) and 0.4(1) respectively.
 \end{abstract}

\newpage
\section{Introduction}
 \label{intro}
 Lattice QCD has had considerable success in the understanding of the
energies of few-quark systems. However, the spatial distributions of the
quarks in these systems have received much less attention. The reasons
for this are two-fold. Firstly, unlike energies, these distributions are
not directly observable, but only arise in integrated forms such as 
sum-rules, form-factors, transition rates etc. Secondly, as will be seen
later, their measurement on a lattice is more difficult and less
accurate than the corresponding energies. In spite of this, it is of
interest to extract lattice estimates of various spatial distributions
and this is the aim of the present study.

There have been earlier lattice measurements of radial distributions.
However, they differ from the present work in several ways.  For
example, in Ref.~\cite{Divitiis} the authors are interested in the
coupling between $B$-mesons and the  $\pi$, which involves a
pseudo-vector coupling of the $\pi$ to a single quark.  In contrast,
here it is the charge and matter distributions that are studied and
these involve vector and scalar couplings. Scalar coupling to single
quarks was also studied in Refs.~\cite{foster},\cite{Fosterthesis}. 
However, there only the scalar sum-rule was evaluated, since that was
primary a study of the dependence of the meson mass on the quark masses.

A knowledge of spatial distributions can be utilized in a variety of
ways. For example, the charge distribution of the light quark
$(\bar{q})$ in a heavy-light meson $(Q\bar{q})$ can be used to check
possible potential models, where the distributions are calculated from 
 wavefunctions generated by some differential equation containing an
interquark potential $V$. In such models \cite{models} there are several
uncertainties -- the form of $V$, the form of the differential equation,
how to include relativistic effects -- some of which are tuned to ensure
the model reproduces the correct  $(Q\bar{q})$-energies. The latter can
be in several partial waves $S_{1/2}, \ P_{1/2}, \ P_{3/2}, \ ...$ and
can be either the observed energies of, for example, the $B$-meson or
the results of a lattice calculation. However, if also spatial
distributions are known a priori from, say, a lattice calculation, then
the uncertainties in such potential models will be reduced. Another way
that the $(Q\bar{q})$ charge distribution could be of use is in the
understanding of multi-quark systems. In few-nucleon systems  (e.g.
$^3He, \ ^4He$)  it is found that the nucleon-nucleon correlations are,
at short distances, very similar to that in the {\em two}-nucleon system
-- with differences arising at large distances due to the different
binding energies \cite{Gl}. This important observation can then be
exploited in models of multi-nucleon systems,   by assuming that the
internucleon  correlations are dominated by their two-nucleon
counterparts. In the corresponding multi-quark case it would be of
interest if a similar simplification were to arise. Therefore, as a
first step in that direction, a knowledge of the two-quark radial
correlation in the basic $(Q\bar{q})$ case is needed -- later to be
compared with those in, say, the $(Q^2 q)$ or  $(Q^2\bar{q}^2)$ systems.
 This probing of the $(Q\bar{q})$ or $(Q^2\bar{q}^2)$ structures could
be carried to a more fundamental level by measuring the form of the
underlying color fields that lead to the interquark potential $V$ and
the radial correlations. Such an extension would be analogous to the
study of these fields in the  $(Q\bar{Q})$ and $(Q^2\bar{Q}^2)$ systems 
-- see Refs.~\cite{fields}.

It should be added that the heavy-light  system $(Q\bar{q})$ is the
quark model equivalent to the hydrogen atom $(Pe^-)$. Therefore, from a
general point of view, it is of interest in any discussions comparing
the properties of two-body sytems constructed from two particles one of
which is very much heavier than the other. Also the interactions in the
two cases have common features -- the coulomb potential $\propto e^2/r$
of the hydrogen atom versus the one-gluon exchange $\propto \alpha /r$
in the heavy-light meson.

In section~\ref{MVR} the Maximal Variance Reduction method, crucial for
extracting reliable results, is briefly discussed.  In
section~\ref{formalism} the formalism is introduced for evaluating the
two- and three-point correlation functions $C(2)$ and $C(3)$.  In
section~\ref{ana}  variational methods for analysing the lattice data
are explained. In section~\ref{results} the results are given and in
section~\ref{con} some conclusions are made. 

 \section{Maximum Variance Reduction method. } \label{MVR}
 
It has been demonstrated  that light-quark propagators can be
constructed in an efficient way using the so-called Maximum Variance
Reduction(MVR) method.  Since this has been explained in detail
elsewhere, for example in Ref.~\cite{MP98}, the emphasis here will be
mainly on the differences that arise when estimating on a lattice the
correlation functions $C(2), \ C(3)$ needed for measuring spatial 
charge and matter densities. In the MVR method the inverse of a positive
definite matrix $A$ is expressed in the form of a Monte Carlo
integration
 \begin{equation}
\label{Gauss1}
A^{-1}_{ji}=
\frac{1}{Z}\int D\phi \ \phi^*_i \phi_j \ \exp(-\frac{1}{2}\phi^*A\phi), 
\end{equation}
 where the scalar fields $\phi$ are pseudo-fermions located on lattice  
sites $i,j$. For a given gauge configuration on this lattice, $N$
independent samples of the $\phi$ fields can be constructed by Monte
Carlo techniques resulting in a stochastic estimate of $A^{-1}_{ji}$ as
an average of these $N$ samples  i.e. $A^{-1}_{ji}=\langle \phi^*_i
\phi_j \rangle$. The $N$ samples of the $\phi$ fields can be calculated
separately and stored for use in any problem involving light quarks with
the same gauge configurations. 

In QCD the matrix of interest is the Wilson-Dirac matrix $Q=1-\kappa M$,
which is not positive definite for those values of the hopping parameter
$\kappa$ that are of interest. Therefore, we must deal with 
$A=Q^{\dagger}Q$, which is positive definite. Since $M$ contains only
nearest neighbour interactions, $A$ -- with at most next-to-nearest
neighbour interactions -- is still sufficiently local for effective
updating schemes to be implemented. In this case the light-quark
propagator   from  site $i$ to site $j$ is expressed as
 \begin{equation}
\label{Gauss2}
G_q=G_{ji}= Q^{-1}_{ji}=\langle (Q_{ik}\phi_k)^*\phi_j\rangle=
\langle \psi^*_i\phi_j\rangle. 
\end{equation}
 This is the key element in the following formalism. 
 The Wilson-Dirac matrix also leads to an alternative form for the above
light-quark propagator from site $i$ to near  site $j$
 \begin{equation}
\label{Gauss3}
G_q'= G'_{ji}=\gamma_5<(Q_{jk}\phi_k)\phi_i^*>\gamma_5=
\gamma_5<\psi_j\phi_i^*>\gamma_5. 
\end{equation}
 Later, it will be essential to use at some lattice sites operators that
are  {\em purely local}. This then restricts us to using at such sites
only  the $\phi$ fields that are located on single lattice sites.  In
contrast the $\psi_i$ fields -- defined as $Q_{ik}\phi_k$ -- are not
purely local, since they contain $\phi$ fields on next-to-nearest
neighbour sites. 

In the above, the term ``Maximal Variance Reduction'' comes from the
technique applied to  reduce the statistical noise in Eq.~\ref{Gauss2}.
The lattice is divided into  two boxes ($0 < t < T/2$ and $T/2 < t < T$)
whose boundary is kept fixed.  Variance of the pseudofermionic fields is
then reduced by numerically solving  the equation of motion inside each
box. This allows the variance of propagators from one box to the other
to be greatly reduced. However, in the case of a three-point correlation
two  propagators are needed and this is best treated by choosing one of 
the points to be on the boundary of the boxes while the other two are
inside  their own boxes. Furthermore, the field at the boundary must be
local to avoid the two propagators interfering with each other. This
means that only the $\phi$ fields can be used on the boundary and there
they couple to the charge or matter operator. For the points in the
boxes,  the temporal distance from the  boundary should be approximately
equal to give the  propagators  a similar degree of  statistical
variance.

\section{The correlation functions $C(2)$ and $C(3,r)$}
\label{formalism}
 In this section an expression is given for evaluating the  two-point
correlation $C(2)$ -- needed for extracting the basic $Q\bar{q}$
energies ($m_{\alpha}$) and eigenfunctions($v^{\alpha}_i$). This is then
extended to the three-point correlation $C(3)$ in order to measure the
radial correlations. Details are given in the Appendix.

\subsection{Two-point correlation functions $C(2)$}
\begin{figure}[ht]
\includegraphics{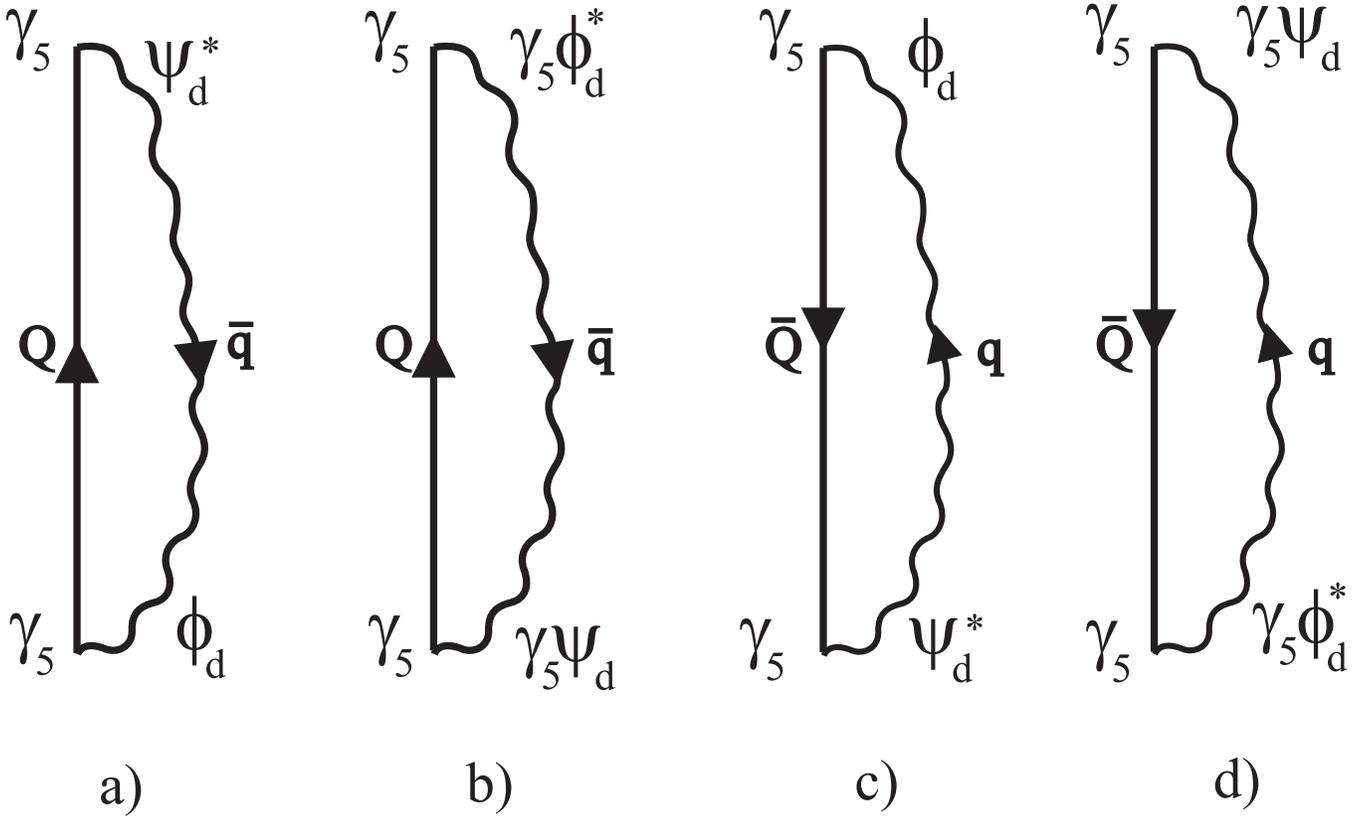}
\caption{The four contributions to the two-point correlation function 
$C(2)$}
\label{fig2e} 
\end{figure}

 Given the above light-quark propagators, then the two-point correlation
functions $C(2)$ needed for extracting the energies of a heavy-light
meson can be expressed in the four ways shown in Fig.~\ref{fig2e}. 
These are all the same upto statistical errors, but their combination
improves the overall measurement. In Figs.~\ref{fig2e} a) and b), the
heavy(static)-quark propagator from site $({\bf x}, \ t)$ to site $({\bf
x'}, \ t+T)$ is simply
 \begin{equation}
\label{GQ}
G_Q({\bf x}, \ t \ ; \ {\bf x'}, \ t+T)=
\frac{1}{2}(1+\gamma_4) U^Q({\bf x},t,T)\delta_{{\bf x},{\bf x'}} ,
 \end{equation}
 where $U^Q({\bf x},t,T)=\prod^{T-1}_{i=0} U_4({\bf x},t+i)$ is the
gauge link product in the time direction.
As shown in the Appendix this leads to the expression
 \begin{equation}
\label{C_2}
C(2,T)=2 \langle  
Re\left[U^Q \ [\sum_{e=3,4} \psi^*_e({\bf x},t+T)\phi_e({\bf x},t)+
\sum_{d=1,2}\phi^*_d({\bf x},t+T)\psi_d({\bf x},t)   ]\right] \rangle,
 \end{equation}
 where   $d$ and $e$ are  Dirac spin indices.

 \subsection{Three-point correlation functions $C(3)$}
 The construction of the three-point correlation functions $C(3)$
follows that of $C(2)$ and are depicted in Fig.~\ref{fig2r}. Since a
probe is now inserted at a distance $r$ from the heavy-quark ($Q$), two 
light-quark propagators enter -- one from $Q$ to the probe and a second
from the probe back to the $Q$. The purpose of the probe is to measure
the charge or matter distribution at a definite point ${\bf r}$.
Therefore, only those light-quark propagators that involve the local
basic field $\phi$ at ${\bf r}$ can be used, since the  $\psi$ field
contains contributions from $\phi$ fields at next-to-nearest neighbour
sites and so is  {\em non-local}. Later, when fuzzing is introduced
similar restrictions will enter. In this work two probes are studied: 
 i) $\Theta ({\bf r})=\gamma_4$ that measures the charge distribution
(actually  the light quark charge normalised to 1) and
 ii) $\Theta ({\bf r}) =1$ that measures the matter distribution of 
the light quark.

As shown in the Appendix, the same techniques introduced to evaluate
$C(2)$ can be extended to $C(3)$  giving
the overall three-point correlation function as
 \begin{equation}
 \label{C_3tot}
 C(3,-t_2, \ t_1, \ {\bf r})=C^Q(3)\mp C^{\bar{Q}}(3)=2\langle  \
X^q_RY^Q_R - X^q_IY^Q_I\ \rangle,
 \end{equation}
 where the $X^q$ and $Y^Q$ can be expressed in terms of  $\phi \phi$ and
$\psi U \psi$ respectively. In Eq.\ref{C_3tot} the relative sign enters
for  $\Theta ({\bf r})=\gamma_4$, since the $q$ and $\bar{q}$ have
opposite charges.

\section{ Analysis}
\label{ana}
 The correlations of interest are essentially obtained from the ratio
$\langle C(3,T)\rangle/\langle C(2,T)\rangle$ by projecting out the ground state
expectation value. However, the latter is only achieved  in the limit
$T\rightarrow \infty.$ In practice, on a given lattice at the maximum 
possible values of $T$, the signal to noise ratio becomes large and
effects from excited states are present. In order to reduce this
contamination, a set of wavefunctions is constructed by fuzzing the
original local wavefunction. 
 These wavefunctions generate a better hadron operator where  the
$Q\bar{q}$-meson is created and destroyed. Then, together with the
original local form, they serve as a  variational basis for analysing
the data.   

\subsection{The effect of fuzzing}
\label{EofF}
Fuzzing enters in two ways. 
\begin{itemize}
\begin{enumerate}

\item Firstly, the basic links containing the gauge field have two
fuzzings. In the standard notation of, for example Ref.~\cite{GMP93}, 
Fuzz1 has 2 iterations and Fuzz2 6 iterations. In both cases, the factor
multiplying the basic link is $f_p=2.5$

i.e. [A fuzzed link] = $f_p\cdot $[Straight link]+ [Sum of 4 spatial
U-bends].

\item The pseudo-fermion field $\phi$ at a given lattice site ${\bf r}$ 
is considered to have three forms: 

\begin{itemize}
\begin{enumerate}

\item The basic form that is simply a function of the single lattice
site   ${\bf r}$. 

\item The field at  ${\bf r}$ is an average of the  fields on the 
neighbouring six lattice sites($i$) i.e. \[ \phi_1 ({\bf r})=\sum_i
U(Fuzz1, {\bf r},{\bf r_i} ) \phi ({\bf r_i}).\]

\item The field at  ${\bf r}$ is an average of the  fields on the 
 six lattice sites($j$)that are {\em two} lattice spacings from ${\bf
r}$ i.e. \[ \phi_2 ({\bf r})=\sum_j U(Fuzz2, {\bf r},{\bf r_j} ) \phi
({\bf r_j}).\]
 \end{enumerate}
\end{itemize}
\end{enumerate}
\end{itemize}
 Therefore, only the basic field is local(L) -- with $\phi_{N=1}$ and  
$\phi_{N=2}$
being increasingly non-local. This means that in the calculation of the
above three-point correlation functions, only the basic $\phi$ field
should be used at ${\bf r}$  -- the position of the probe insertion. 
 There are now two reasons for this: 1) The field on the boundary at
$t=0$ must be local. 2) The operator insertion must be local.  This
restriction does not occur elsewhere, so that the $\psi$ fields, which
connect directly to the heavy quark, can be the fuzzed forms constructed
from $Q_{ik}\phi_N$. This means that the two- and three- point
correlation functions have the same size ($3\times 3$) of overall 
correlation matrix -- $LL, \  LF_1, \ LF_2,  \ F_1F_1,    \ F_1F_2$ and
$F_2F_2$.

\subsection{The variational method}
\label{variational}

There are many ways of analysing the above correlations in order to
extract the quantities of interest i.e. energies and wavefunctions.
Here a variational method  described in Ref.~\cite{McN+M} is applied. 

Firstly, the two-point correlation data $C(2)$  are analysed to give the
energies ($m_{\alpha})$ and eigenvectors $({\bf v})$ for the 
$Q\bar{q}$-system. These are then used in analysing the three-point
correlation  data $C(3)$ to give the charge and matter densities. 

Consider the  correlation function $C(2,T)$  as an $n*n$ matrix -- upto
3*3 in this case with the elements  $LL, \ LF_1, \ .... F_2F_2$. Each
element $C_{ij}(2,T)$ is then fitted with the form
 \begin{equation}
\label{Cij}
C_{ij}(2,T)\approx \tilde{C}_{ij}(2,T)=\sum_{\alpha =1}^{M}v_i^{\alpha}
\exp(-m_{\alpha}T)v_j^{\alpha},
\end{equation}
 where  $m_1$ is the ground state energy of the heavy-light meson. The 
statistically independent matrix elements of $C(2)$ are then fitted by
varying the  $v_{i,j}^{\alpha}$ and $m_{\alpha}$ to minimise the
difference between $C(2)$ and  $\tilde{C}(2)$.

We illustrate the procedure for the 2*2 case, where the $C(2)$ can be
expressed as the product of three 2*2 matrices
 \begin{equation}
\label{Cmat}
C(2)=c^T \ \left( \begin{array} {cc}
\exp(-m_1 \ T)&0\\
0&\exp(-m_2 \ T)     \end{array}\right) \ c \ , \ {\rm where}  \ 
c=\left( \begin{array} {ll}
v^1_{L}&v^1_{F1}\\
v^2_{L}&v^2_{F1}     \end{array}\right)
 \end{equation}
 and $c^T$ is the transpose of $c$. In Ref.~\cite{McN+M} the rows of $c$
are called the ${\bf v}$-vectors $v_i^{\alpha}$. Once the $c$ matrix is
known, any 2*2 correlation matrix $C$ can be evaluated for the
ground(excited) state wavefunction corresponding to the extracted
eigenvalue $m_1(m_2)$ by reversing the above procedure to give
 \begin{equation}
\label{Cin}
\bar{C}= (c^T)^{-1} \ C \ c^{-1} \ \ {\rm i.e.} \ \ 
\bar{C}_{\alpha \beta}=u^{\alpha}_i \ C_{ij}(T) \ u^{\beta}_j,
\end{equation}
 where the $u^{\alpha}_i$ are components of the ${\bf u}$-vectors in 
Ref.~\cite{McN+M}. These ${\bf u}$-vectors are the columns of the $
c^{-1}$ matrix and satisfy the condition 
 \begin{equation}
\label{orth}
v^{\alpha}_i \ u^{\beta}_i= \delta_{\alpha \beta}. 
 \end{equation}
 For the 2*2 case 
\begin{equation}
 \label{ui}
 {\bf u}^1=[v^2_{F1}, \ -v^2_L]/det(c), \ \ \ {\bf u}^2=[-v^1_{F1}, \ \
v^1_L]/det(c).
 \end{equation}
 In the case where $C$ is the above two-point correlation function
$C(2)$ and $\tilde{C}$ in Eq.~\ref{Cij} is a perfect fit to $C$, then
the operation in Eq.~\ref{Cin} would result in $\bar{C}$ being diagonal
with the diagonal elements simply being $\exp(-m_{\alpha}a)$. Of course,
in practice, the fit is never perfect and  the off-diagonal elements of
$\bar{C}$ are a measure of this goodness of fit. This will be
demonstrated later. However, as pointed out in Ref.~\cite{McN+M}, for
other correlation functions there is no reason for $\bar{C}$ to be
diagonal.

\section{Results}
\label{results}
 The results are presented in two distinct parts. Firstly, the two-point
correlation function is analysed to give the ground and excited state
energies and eigenvectors i.e.  $m_1$, $m_2$ in Eq.~\ref{Cmat} and ${\bf
u}^1$, ${\bf u}^2$ in Eq.~\ref{Cin}. Secondly, these eigenvectors are
used to extract the charge and matter radial distributions from the
three-point correlation functions.

The actual pure gauge configurations (20 in number) and the
pseudofermion fields $\phi$ (24 per gauge configuration) were taken from
the tabulation generated for the work of Ref.~\cite{MP99}. These are for
a $16^3\times 24$ lattice  with $\beta=5.7$ with  the
Sheikholeslami-Wohlert improved clover action with $c_{SW}=1.57$ --
corresponding to a lattice spacing of $a\approx 0.17$fm -- and a hopping
parameter $\kappa=$0.14077. The latter corresponds roughly to the
strange quark mass. This can be seen from ref\cite{shan} where the same
parameters in the light-light   system $(q\bar{q})$ predict a vector
meson to pseudoscalar meson mass ratio corresponding to strange quarks.
More details can be found in Ref.~\cite{MP98}.

\subsection{Analysis of the two-point correlation function $C(2,T)$}
 Essentially the energies $(E)$, in lattice units,  are extracted from
the $C(2,T)$ in Eq.~\ref{C_2} using
 \begin{equation}
 \label{E_2} 
 E[C(2,T)]=- \ln \ [\frac{\langle C(2,T)\rangle}{\langle
C(2,T-1)\rangle}] \  \ \ {\rm as} \  T\rightarrow \infty.
 \end{equation}
 In Fig.~\ref{fig3abc}a) the basic $ C(2,T)$ are plotted for the three 
diagonal matrix elements where the fields are: i) purely local(L),  ii)
all Fuzz1,  iii) all Fuzz2 -- in the notation of subsection \ref{EofF}.
They are all seen to be well determined with the errors only being
significant for Fuzz2 with $T>10$. In Fig.~\ref{fig3abc}b) the results
$E[C_{ii}(2,T)]$ from Eq.~\ref{E_2}  are plotted for the three sets of
diagonal matrix elements with $i=L, \ F1, \ F2.$ There it is seen that
only $E[C_{F1F1}(2,T)]$ shows a clean plateau that extends from $T=5$ to
9 with a value about 0.88(1).

To combine these results by a variational calculation using
Eq.~\ref{Cij} two numbers are fixed: 

i) $M$ -- the number of
energies being included. Here, this is taken to be the same as the
number of paths for each energy and results in the correlation
matrices  being square.

ii) $T_{min}$ -- the minimum value of $T$ used in the fit.

Here we consider four possibilities to check the dependence of the final
results on this fitting procedure:

Case 1): In Eq.~\ref{Cij}, the $ \tilde{C}_{ij}(T)$ are defined in terms
of 3 paths (i.e. $i,j=1,2,3$) and 3 exponentials (i.e. $M=3$) with
$T_{min}=3$.  This includes the Local and both Fuzz1 and Fuzz2 paths.

Case 2): Same as Case 1) but with 2 paths and 2 exponentials.
This  includes only the Local and Fuzz1 paths.

Cases 3,4) are the same as Cases 1,2) but with $T_{min}=4$. 

Minimising the difference between the $C(2,T)$ and  $\tilde{C}(2,T)$ in
Eq.~\ref{Cij} gives the parameters in Table~\ref{Tablecs}. These are
surprisingly good fits, when it is realised that the errors on the
$C(2,T)$ are, in most cases, much less than $1\%.$ However, only Case 3
gives $\chi^2/n_{dof}(2)<1$ and so this is the one that will be used in
most of this study.

In Fig.~\ref{fig3abc}c) the  results for $E[\bar{C}_{\alpha \alpha}]$ 
from Eq.~\ref{E_2} are plotted for the two sets of  diagonal matrix
elements $\bar{C}_{11}(2,T)$ and $\bar{C}_{22}(2,T)$ for Cases 3 and 4.
As expected, $E[\bar{C}_{11}(2, T)]\approx 0.86(2)$ and  $
E[\bar{C}_{22}(2,T)] \approx 1.24(5)$ -- energies that   are consistent
with the values of $m_1$ and $m_2$ in  Table~\ref{Tablecs}. As a
check on the off-diagonal matrix element $\bar{C}_{12}(2,T)$, we
evaluate
 \begin{equation}
\label{Re2} 
 R[\bar{C}_{12}(2,T)]=\frac{\langle \bar{C}_{12}(2,T)\rangle}{\langle
\bar{C}_{11}(2,T)\rangle} \ . 
 \end{equation}
 This is seen to be at the $1\%$ level. The conclusion to be drawn from
Fig.~\ref{fig3abc}c) is that $\bar{C}$ is, indeed, approximately diagonal
with $\bar{C}_{\alpha \beta}(2,T)\approx \exp(-m_{\alpha}T)
\delta_{\alpha \beta}.$ These results will serve as a comparison
when analysing $C(3,T,r)$ later.

\subsection{Analysis of the three-point correlation function for sum
rules}
 The charge and matter radial distributions $F(r, \Theta)$ of  
the light quark in the  $Q\bar{q}$ system are extracted using 
 \begin{equation}
\label{Fr}
 F[C(\Theta, T,r)]=\frac{\langle C(3,\Theta,T,r)\rangle}{\langle
C(2,T)\rangle},
 \end{equation}
 where $ \Theta=\gamma_4$ or 1.
However, before showing these radial distributions, it is of interest
to first study the corresponding sum rules
 \begin{equation}
\label{sumr}
  F^{sum}[C(\Theta, T)]=\frac{\sum_{{\bf r}}\langle C(3,\Theta,T,{\bf
r})\rangle}{\langle C(2,T)\rangle} = 
\frac{\langle C^{sum}(3,\Theta,T)\rangle}{\langle C(2,T)\rangle},
 \end{equation} where  $\sum_{{\bf r}}$ represents the sum over all
spacial lattice sites at time $t=0$ -- the time when the probe acts. The
summation  can be easily carried out exactly on the lattice. For the
charge distribution this sum rule should, {\em in the continuum limit}, 
simply yield the charge of the light quark, which we have chosen to  
normalise to unity. For
the matter distribution the  situation is less clear -- see Refs.
~\cite{foster},\cite{Fosterthesis}. 

In Fig.~\ref{Fig.4}a) we show  $ F^{sum}[C_{ii}(\gamma_4, T)]$ for the
three cases $i=L, \ F1, \ F2$. There it is seen that, at large $T$,  the
sum rule for $i=L$ becomes $1.05(5)$ and so is consistent with unity.
However,  the results for $i=F1, \ F2$ appear to be   somewhat greater
than unity for $T> 5$ converging to $1.25(5)$ and $1.4(1)$.
  When these results are combined using the $\bf{u}$-vectors in 
Table~\ref{Tablecs},  then -- as seen in Fig.~\ref{Fig.4}(b) -- the
ground state sums   $ F^{sum}[\bar{C}_{11}(\gamma_4, T)]$ for both Cases
3 and 4 are dominated by the $ F^{sum}[C_{F1F1}(\gamma_4, T)]$.
Consequently, they are again distinctly greater than unity with  both
Cases 3 and 4 tending to about  $1.30(5)$. This seems unavoidable since,
as seen  from Table~\ref{Tablecs}, the $\bf{u}$ vector components
$u^1_{F1}$ and $u^1_{F2}$ are an order of magnitude larger than $u^1_L$.
 In Fig.~\ref{Fig.4}b) the sum-rule for the first excited state  $
F^{sum}[\bar{C}_{22}(\gamma_4, T)]$ is also shown. This appears to be
approaching unity at about $T\approx$7 -- 8. However, the signal is
swamped by the error bars at larger $T$.

 The correlations  $C^{sum}_{ij}(3)$ are not so
well diagonalised as the $C_{ij}(2)$ were forced to be earlier.
A measure of this is plotted in Fig.~\ref{Fig.4}b) as 
 \begin{equation}
\label{sumr0}
  R^{sum}[\bar{C}_{12}(\Theta, T)]= 
\frac{\langle \bar{C}^{sum}_{12}(3,\Theta,T)\rangle}
     {\langle \bar{C}^{sum}_{11}(3,\Theta,T)\rangle }.
 \end{equation}
 This is seen to be at the $10\%$ level -- an order of magnitude larger
than the corresponding $R[\bar{C}_{12}(2,T)]$ in Eq.~\ref{Re2}.
Therefore, it is seen that  the $\gamma_4$ sum rule calculated in the
above way has two undesirable features:

\begin{itemize}
\begin{enumerate}
 \item The ground state sum rule $ F^{sum}[\bar{C}_{11}(\gamma_4, T)]$
is not consistent with unity for large $T$ being more like $1.30(5)$.
Only in the continuum limit should the sum be unity. 
 To some extent this must be expected, since in the present
non-continuum situation  the lattice vector current is not conserved. In
principle this can be corrected  in various ways. Unfortunately,  at the
present  value of  $\beta=5.7$, perturbative expressions would exhibit
poor convergenge and so be unreliable.   Non-perturbative corrections
are discussed in the recent review  in  Ref.~\cite{bowler}. There an
improved vector operator is introduced and calculations  performed in
the quenched approximation at $\beta=$ 6.2 and 6.0.  However, they find
that, whereas the $\beta=$6.2 results are satisfactory, those at
$\beta=$ 6.0 are not -- suggesting that $O(a^2)$ discretisation errors
are not small at the larger lattice spacing. The situation would be even
worse at the value of $\beta=5.7$ used here. 
 Even so, it is illustrative to see  the results at the higher
$\beta$'s, since they may indicate what could possibly be expected at
lower $\beta$'s.  In Ref.~\cite{bowler}, when the expectation value of
the renormalised  vector current is expressed as
 \begin{equation}
\label{Bow}
\langle J^R \rangle =F^V \langle \gamma_4(x) \rangle,
 \end{equation}
 where $\langle\gamma_4(x) \rangle$ is the quantity evaluated in
Eq.~\ref{sumr}, they find that  $F^V\approx 0.8$. Similar reductions 
are found in Ref.~\cite{Divitiis} for the axial vector operator. Of
course, the above argument could be reversed  to say that the present
charge sum-rule measurement of 1.30(5) gives a non-perturbative estimate
of $F^V$ as 0.77(3). Later, when individual radial contributions are
extracted as in Eq.~\ref{Fr}, this continuum effect could be
incorporated approximately by the renormalisation
 \begin{equation}
\label{renorm}
 F[\bar{C}_{\alpha\alpha}(\gamma_4, T,r)]\rightarrow 
\frac{F[\bar{C}_{\alpha\alpha}(\gamma_4, T,r)]}
{F^{sum}[\bar{C}_{\alpha\alpha}(\gamma_4, T)]}.
\end{equation}

\item The off-diagonal terms, as illustrated by  $ 
R^{sum}[\bar{C}_{12}(\gamma_4, T)]$ in Eq.~\ref{sumr0}, are not zero.
This is potentially more disturbing than point 1), since we wish to
identify $ F[\bar{C}_{11}(\gamma_4, T,r)]$ as the ground state charge 
distribution, and so any off-diagonal effects would be difficult to
interpret. Also they cannot be easily renormalised away as in
Eq.~\ref{renorm} .
 \end{enumerate}
\end{itemize}

So far the basic wavefunctions $u_i^{\alpha}$ have been determined, from
Eq.~\ref{Cij}, via the $v_i^{\alpha}$ i.e. by considering only the
two-point correlations $C(2)$. In an attempt to overcome the two above
problems with $ F^{sum}[\bar{C}(\gamma_4, T)]$, the analysis of the data
is now carried out not only by fitting $C(2)$ but also some features of
$C(3)$.
 \begin{itemize}
\begin{enumerate}
\item In analogy to Eq.~\ref{Cij} the sum rule data are fitted with the
 expression    
\begin{equation}
 \label{srfit}
C^{sum}_{ij}(3,\gamma_4,T)\approx \tilde{C}^{sum}_{ij}(3,\gamma_4,T)=
\sum_{\alpha =1}^{M}\sum_{\beta =1}^{M}v_i^{\alpha}
\exp[-m_{\alpha}t_1]x^{\alpha\beta}\exp[-m_{\beta}(T-t_1)]v_j^{\beta},
 \end{equation}
 where the $v_i^{\alpha}$ are from Table~\ref{Tablecs}  -- the earlier
result of  minimizing the $C_{ij}(2)$ -- but the $x^{\alpha \beta}$ are
treated as  free parameters. However, a restriction must be made on the
values of $T$ used  in Eq.~\ref{srfit}. From Table~\ref{Tablecs} it is
seen that $T_{min}$ must be at least 4 to ensure $\chi^2(2)/n_{dof}(2)$
is comparable to unity. Therefore, in  $C^{sum}(3)$ each of the two
propagators should be at least 4 euclidean time steps i.e. in
Eq.~\ref{srfit} we should have $t_1 \ge 4$ and $(T-t_1)\ge 4$. This
means that only the $C(3,T)$ data with $T_{min}(3)\ge 8$ should be
included in any fitting procedure. However, in the following a series of
$T_{min}(3)$ values, ranging from 4 to 9, are used to see how the  final
results depend on $T_{min}(3)$. Therefore, in Case 3, i.e.  3*3 with
$T_{min}(3)=4[8]$, this involves  fitting 42[18] pieces of data with 6
parameters and in Case 4, i.e. 2*2 with $T_{min}(3)=4[8]$, this involves
fitting 21[9] pieces of data with 3 parameters.  In 
Table~\ref{Tablexs}, the results for   the $\chi^2/n_{dof}$ are shown
separately for the fits to $C(2)$ and $C^{sum}(3)$. There for Case 3 it
is seen that -- as with  $\chi^2(2)/n_{dof}(2)$ -- the
$\chi^2(3)/n_{dof}(3)$ are also $\approx 1$ provided  $T_{min}(3)\ge 6$.
 This result can be compared directly with  Fig.~\ref{Fig.4}b), since it
is simply an alternative analysis under similar constraints. It is seen
that the values of $x^{11}=1.35(5)$ and $x^{22}\approx 1$ are indeed
consistent with the asymptotic values of $F^{Sum}$ in 
Fig.~\ref{Fig.4}b).  On the other hand, $x^{12}$ cannot be compared
directly with $R^{sum}[\bar{C}_{12}(\gamma_4, T)]$ in Eq.~\ref{sumr0}.
All that can be said there is that both analyses result in non-zero and
negative values for $x^{12}$. 
 \item In an attempt to overcome this last point,  in Cases $3'$ and
$4'$, all three off diagonal terms $x^{\alpha\beta}$  are fixed at zero.
For a given $T_{min}(3)$, this decreases $x^{11}$ but at the expense of
increasing $\chi^2(3)/n_{dof}(3)$. The outcome is that for
$\chi^2(3)/n_{dof}(3)<1$ with  $x^{12}=0$ and $T_{min}(3)=8$ we get 
$x^{11}$ is 1.29(3) -- a number consistent with 1.30(5) from 
Fig.~\ref{Fig.4}b). However, $x^{22}$ changes violently, even though  
the plot corresponding to Fig.~\ref{Fig.4}b) is very similar -- with, in
particular,  $ F^{sum}[\bar{C}_{22}(\gamma_4, T)]$ again approaching
unity and  not zero as would be expected from the value for $x^{22}$ in 
Table~\ref{Tablexs}. To check this last unexpected result, the analysis
program was run using the {\em model} results
$\tilde{C}^{sum}_{ij}(3,\gamma_4,T)$ instead of the lattice data
$C^{sum}_{ij}(3,\gamma_4,T)$. The plot corresponding to
Fig.~\ref{Fig.4}b) now has for the  excited state  $
F^{sum}[\bar{C}_{22}(\gamma_4, T)]\approx 0.005$ for all values of $T$
-- consistent with Table~\ref{Tablexs}. This difference with $x^{22}$
demonstrates the need to try to improve the  $\chi^2(3)/n_{dof}(3)$ at
smaller values of $T_{min}(3)$.  It appears  that the results for the
excited states are very dependent on the ${\bf v}$-vectors, since they
involve delicate cancellations. Of course, the main concern is the
consistency between  $x^{11}$ and $F^{sum}[\bar{C}_{11}(\gamma_4, T)]$
and this emerges intact.
 \item In the above, the data for $C_{ij}(2,T)$ and
$C^{sum}_{ij}(3,\gamma_4,T)$  were fitted separately. Therefore, now a
combined fit is made  using  Eqs.~\ref{Cij} and \ref{srfit}  i.e. for 
Case 3(3$'$) with $T_{min}=4$,  $T_{min}(3)=8$, this  involves fitting
54+18 pieces of data for $C(2)+C(3)$ with 12+6(3) parameters. However,
this has a completely negligible effect e.g. in Case 3, $T_{min}=4$, 
$T_{min}(3)=8$ the values of  $\chi^2(2)/n_{dof}(2), \
\chi^2(3)/n_{dof}(3)$ and $\chi^2(2+3)/n_{dof}(2+3)$ change from 0.647,
0.718, 0.673 to 0.652, 0.699, 0.669 and $x^{11}$ from 1.3274(249) to
1.3274(255).  The conclusion is that for the sum rules there is no
benefit in fine tuning the results by simultaneously fitting $C(2)$ and
$C(3)$.
 \end{enumerate}
\end{itemize}
\subsection{Analysis of the three-point correlation function 
for radial distributions}
 In Figs.~\ref{Fig.5}a) and \ref{Fig.5}b) are shown, for
$r/a=0,\ldots,5$,  the ratios 
 \begin{equation}
\label{Fcr}
 F[\bar{C}_{11}(\gamma_4, T,r)]=
\frac{\langle \bar{C}_{11}(3,\Theta,T,r)\rangle}
{\langle \bar{C}_{11}(2,T)\rangle},
\end{equation}
 based on Eq.~\ref{Fr} using the ${\bf v}$-vectors of Case 3.  These all
exhibit, to a certain extent, a plateau as $T$ becomes large. However,
for  $r/a>5$  the error bars become very large beyond $T=8$. No attempt
will be made at this stage to extract the asymptotic values giving the
radial distributions.

The second procedure for analysing the radial distribution three-point
correlation functions $C(3,\theta,T,r)$ is similar to the one 
followed above but using the expression
 \begin{equation}
\label{C3fit}
C_{ij}(3,\gamma_4,T,r)\approx \tilde{C}_{ij}(3,\gamma_4,T,r)=
\sum_{\alpha =1}^{M}\sum_{\beta =1}^{M}v_i^{\alpha}
\exp[-m_{\alpha}t_1]x^{\alpha\beta}(r)\exp[-m_{\beta}(T-t_1)]v_j^{\beta},
 \end{equation}  
 analogous to Eq.~\ref{srfit}  for the sum-rule but with all the 
$r$-dependence being put into the $x^{\alpha\beta}(r)$.
Two types of fit are made:
 \begin{itemize}
 \begin{enumerate}
 \item The ${\bf v}$-vectors,  obtained by minimising the $C(2)$,  are
used in Eq.~\ref{C3fit} and for each value $r$ the $x^{\alpha \beta}(r)$
are varied to ensure a good fit to $C_{ij}(3,\gamma_4,T,r)$ by the model
form $\tilde{C}_{ij}(3,\gamma_4,T,r)$.
 \item 
 A simultaneous fit of $C_{ij}(2,T)$ and $C_{ij}(3,\gamma_4,T,r)$ using 
$\tilde{C}_{ij}(2,T)$ and $\tilde{C}_{ij}(3,\gamma_4,T,r)$ of  
Eqs.~\ref{Cij} and ~\ref{C3fit}.  
 \end{enumerate}
\end{itemize}
 To achieve a $\chi^2(2+3)/n_{dof}(2+3)\approx 1$ is now more difficult 
than the earlier $C(2), \ C^{sum}(3)$ fit, since the
$C_{ij}(3,\gamma_4,T,r)$ have error bars that are  smaller than the
corresponding sum rule correlations. For example, the most important
correlations $i=j=F1$ have the values $C(2,T=4)=0.0752(3)$ and
$C(3,T=8,r/a=2)=0.0000528(9)$ compared with $C^{sum}(3, T=8)=0.029(1)$
i.e. the errors on $C(2)$ are $\approx 0.5\%$,  those on $C^{sum}(3)$
are $\approx 5\%$, but   those on $C(3,r)$ are only $2\%$. This is seen
in Table~\ref{Table.xr33p}, where  in particular for $r/a=1$  the values
of  $\chi^2(3)/n_{dof}(3)$ are all greater than unity. In this table it
is also seen that the  radial distribution of the  ground state charge
density, $x^{11}(r)$,  is well determined and is consistent with the
plateaux in Figs.~\ref{Fig.5}a) and b). When the $x^{11}(r)$ are plotted
on a semi-log scale as in  Fig.~\ref{Fig6cm.}, the distribution for
$r/a \le 4$ is  approximately a  straight line suggesting that
$w_c(r)=A_c\exp(-r/r_c)$, where $w_c^2(r)=x^{11}(r)$. The function
$w_c(r)$ is then  interpreted as a radial wavefunction.  However, the
other diagonal matrix element, $x^{22}(r)$, is much less well determined
-- as is seen in the lower half of  Table~\ref{Table.xr33p}. In fact,
for some values of $r$, it appears to be  slightly negative. But the
actual values are  very small and could well be consistent with zero.
Only for Case 3 at $r/a=0$ is there a definite signal with
$x^{22}(0)\approx 0.19$. This suggests $x^{22}(r)$ is very sharply
peaked compared with $x^{11}(r)$. However, it must be remembered that in
Case 3 the off-diagonal terms $x^{\alpha \beta}(r)$ are not forced to be
zero and so the interpretation of $x^{22}(r)$ is not clear. In Case 3$'$
where the off-diagonal terms $x^{\alpha \beta}(r)$ are  forced to be
zero, no signal can be extracted at $r=0$. There is certainly no sign of
a node that would be expected of an excited s-wave.

Fig.~\ref{Fig6cm.} also suggests that for $r/a \ge 4$ the function
$w(r)$ drops off faster than the above simple exponential. Such an
effect is expected at sufficiently large $r$ when the linear rising
confining potential becomes important. Then as $r\rightarrow \infty$ 
the wavefunction should have an $\exp[-(r/r_1)^{3/2}]$ asymptotic form.
Unfortunately, for $r/a >5$ the errors on the data become too large. So
that the above observation rests completely on the $r/a=5$ data, which
itself is rather poor. Even so, there the errors are still sufficiently
small to support this. It is planned to extend the present calculation
to off-axis points. The $r/a=5$ results can then be checked by
performing simulations  at $x/a=3, \ y/a=4$. The resulting data should
be more accurate since each $(x,y)$-pair arises in 24 different ways
--not just 6 as for the on-axis points with $r\not= 0$.

The actual parameters defining $w_c(r)$ can be extracted in a variety of
ways depending on $T_{min}(3)$ and the range of $r$ values used in the
fit.  But as seen in Table~\ref{Table.Ar0a} they appear to be quite
stable for Case 3 and 3$'$ separately. For Case 3,  $A_c\approx 0.23(1)$
and $r_c/a \approx 2.1(1)$ and for Case 3$'$,  $A_c\approx 0.26(1)$ and 
$r_c/a \approx 1.9(1)$. Given these numbers, then estimates of the
sum-rule are $I_c=\int^{\infty}_0 d{\bf r} w_c^2(r)=\pi A_c^2 r_c^3$.
These are shown in the last column of Table~\ref{Table.Ar0a} and are 
all  $1.6(1)$ for Case 3 and 1.5(1) for Case 3$'$. The corresponding
numbers for the sum rule obtained directly from the lattice and shown in
 Table~\ref{Tablexs} are 1.35(5) and 1.25(5) respectively. This
difference between the two methods is not surprising, since the
integrand in $I_c=\int^{\infty}_0 d{\bf r} w_c^2(r)$ is maximum at
$r/a \approx 2$ and half of the sum $I_c$ is lying inside this value of
$r$. Therefore, lattice artifacts, present at small values of $r$, may
play a role. These are expected to affect $I_c$ more than the direct
results  of Table~\ref{Tablexs} -- a point that can be checked by
estimating $x^{11}$ away from axes e.g. not only at $(\pm 1,0,0), \
(0,\pm 1, 0), \ (0,0,\ \pm 1)$ but also at $(\pm 1,\pm 1,0)$ etc.

In addition to fitting the $x^{11}\approx w_c^2(r)$ with simply the  two
parameter function $w_c(r)=A_c\exp(-r/r_c)$, attempts were made with the
three parameter function $w_c(r)=A_c\exp[-(r/r_c)^p)$. As seen in 
Table~\ref{Table.Ar0a}, this results in values of $p$ that are somewhat
greater than unity -- as expected from Fig.~\ref{Fig6cm.} where the
$r/a=5$ point appears to drop below the earlier simple exponential. This
also has the effect of decreasing the value of $I_c=\int^{\infty}_0
d{\bf r} w_c^2(r)$ from the simple exponential value of $\pi A_c^2
r_c^3$. However, since about $90\%$ of $I_c$ comes from $r$ values less
than 5, this in practice can have only a minor effect.

 Assuming the simple exponential form, an estimate of the mean-square
charge radius is $\langle r^2 \rangle = 3r^2_c$  in lattice units of
$a\approx 0.17$fm. For Cases 3 and 3' this gives 0.38(3) and 0.32(1)
fm$^2$ respectively. 

\subsection{Analysis of the three-point correlation function 
for the Matter radial distributions }
 The previous two subsections have dealt with the charge radial
distribution, where the operator $\Theta$ in Eq.~\ref{Fcr}
is $\gamma_4$ for probing the $\bar{q}$ charge. In this subsection
$\Theta=1$, which probes the $\bar{q}$ matter. 

In Table ~\ref{Tablefxs} the values of the $x^{\alpha \beta}$ in Cases 3
and $3'$ are given. Here it is seen that $x^{11}$ the matter sum rule
for the ground state is more erratic than its charge counterpart in
Table ~\ref{Tablexs} -- a point clearly seen in the corresponding plot
in Fig.~\ref{Fostersum}. A reasonable estimate for $x^{11}$ is 0.4(1).
This value is consistent with the corresponding estimates in 
Refs.~\cite{Fosterthesis} and ~\cite{MP98} and for $12^3\times 24$
lattices. These were made  by  employing data from different hopping
parameters ($\kappa$) and  using the identity 
 \begin{equation}
\label{ident}
x^{11}=\frac{d(am_1)}{d\kappa ^{-1}},
 \end{equation}
 where $am_1$ is the ground state energy and $\kappa$ the hopping 
parameter -- see Ref.~\cite{foster}.  When the $m_1$'s correspond to the
cases where the light meson is of  about one and two strange  quark
masses, Refs.~\cite{Fosterthesis} and ~\cite{MP98} give   0.34(8) and
0.31(6) respectively -- consistent with the present value of 0.4(1).

These larger values are also consistent with the following simple
estimate -- again using the above identity in Eq.~\ref{ident}. If the
$Q\bar{q}$-meson mass $(am_1)$ is taken to be simply the sum of the
quark masses $am_Q+am_q$ and $\kappa ^{-1}=8+2am_q$, then 
 \begin{equation}
\label{ident2}
x^{11}=\frac{d(am_Q+am_q)}{d(8+2am_q)}=0.5.
 \end{equation}
 Another reason for expecting $x^{11}< 1$ also follows from a potential
approach using the Dirac equation. If the $\bar{q}$ is treated as a 
particle in a potential generated by the $Q$, then its  wavefunction
will be of the form $(f,g)$, where $f(g)$ are the large(small)
components of the Dirac wavefunction. The charge of the  $\bar{q}$ will
then be simply $I_C=\int d{\bf r}[f^2(r)+g^2(r)]$, which by the
normalisation will be unity. However, when the charge
operator($\gamma_4$) is replaced by the matter operator (unity), then
the correponding integral is now   $I_M=\int d{\bf r}[f^2(r)-g^2(r)]$.
In the non-relativistic limit $I_C=I_M$. But as relativistic effects
enter i.e. $g^2$ increases from zero, then $I_M$ becomes less than $I_C$
i.e. less than unity. An indication of this is seen in the lattice
results in Table 10.3 in  Ref.~\cite{Fosterthesis}, where $x^{11}$
decreases from 0.21(8) to 0.11(5) as the $\bar{q}$ mass decreases from
about two to one strange quark masses. This also shows -- as expected --
that the interquark potential is more than one-gluon-exchange
$(-\alpha/r)$, since the latter  results in $g/f=(1-\gamma)/\alpha$,
where $\gamma=\sqrt{1-\alpha^2}$ -- a ratio that is independent of the
light quark mass -- see, for  example, \cite{B+D}. For $\alpha=0.3$ this
gives $g/f=0.15$ -- a number that is much smaller than the $\approx
0.75$ suggested by the charge and matter sum-rules measured above on the
lattice.

In Table~\ref{Table.matter} the matter densities are given in analogy
with the charge densities of Table~\ref{Table.xr33p}. Comparison of
these two tables shows that the ground state matter distribution  
$x(matter)=x^{11}_m$  decays faster than the corresponding ground state
charge  distribution $ x(charge)=x^{11}_c$.  At $r=0$ the two are
comparable but, by $r/a=4$,  $x(matter)$ is only $25\%$ of $x(charge)$. 
This is also seen in Figs.~\ref{Fig6cm.}.  Now the $r/a=5$
data is even more uncertain than the  earlier charge data. It is,
therefore, not quoted. 

 Table~\ref{Table.matterAr0} shows the parameters $A_m$ and $r_m$ when
$w_m(r)$ is parametrized as  $w_m(r)=A_m\exp(-r/r_m)$. There it is seen
that $A_c$, for the charge distribution, and  $A_m$ are comparable, but
that $r_m\approx 1.55(5)$ compared with the charge range of $r_c/a
\approx 2.0$. This difference between $r_c$ and $r_m$ means  the
interpretation that $w_c(r)$ and $w_m(r)$ are both $\bar{q}$
wavefunctions is not so direct, since in the most naive models one would
expect $w_c(r)=w_m(r)$.

\section{Conclusion}
\label{con}

In this paper a first study has been made of the radial structure of a
heavy-light meson -- the quark equivalent of the hydrogen atom. This can
be considered as a partial extension of Ref.~\cite{MP98} in which only
the energies of heavy-light mesons were measured on a lattice. Here the
emphasis is on the charge and matter radial distributions. So far these
distributions have only been extracted for the ground state, with the
extension to other partial waves --  as in  Ref.~\cite{MP98} -- only now
beginning to be studied. A further extension would be a study of the
form of the underlying color fields that lead to these radial
correlations. This would be analogous to the studies in 
Refs.~\cite{fields} for the $(Q\bar{Q})$ and $(Q^2\bar{Q}^2)$ systems.

The main result is  in Figure~\ref{Fig6cm.}, where it is seen that both
the charge and matter radial distributions fall off more or less
exponentially as $A^2_i\exp[-2r/r_i]$. The amplitudes $A_i$ are about
the same with $A_c\approx 0.26(1)$ and  $A_m\approx 0.29(1)$, whereas 
the charge range $r_c/a \approx 1.9(1)$ is considerably longer than the
matter range $r_m/a \approx 1.4(1)$. This difference is reflected in the
spatial sum-rule, which is about 1.30(5) for the charge and 0.4(1) for
the matter. The fact that the charge sum-rule is not unity, as would be
expected from vector current conservation, can be attributed to the
finite spacing of the lattice. As shown in Refs.~\cite{Divitiis} and 
~\cite{bowler} this can easily lead to 10-20\% reductions. 

It should be added that there are other definitions of  
$Q\bar{q}$-wavefunctions. Here the relative wavefunction [$w(r)$]  is
assumed to be real with $w^2(r)$ giving the charge density -- the
quantity actually measured from $\langle C(3,T)/C(2,T)\rangle$. Another
form can be extracted directly from a two-point correlation function in
which the operators at the sink and source are of spatial size $r_1$ and
$r_2$ respectively. The ground state correlation can then be fitted with
$w_{BS}(r_1)w_{BS}(r_2)\exp(-m_1T)$ to give a  $w_{BS}(r)$, which can be
identified with the Bethe-Salpeter wavefunction -- see Ref.~\cite{MP98}.

 The above charge and matter radial densities are related  by a simple
fourier transform to the   momentum space vector and scalar form factors
 $[F_{v,s}(q^2)]$ of the $B$-meson. Unfortunately, the present densities
are  not accurate enough over a sufficiently large range of $r$ to 
perform this transform. However, a simple model of these form factors 
is that they are dominated by the pole due to the lightest meson of mass
$m$ i.e. they have a form $\propto (q^2+m^2)^{-1}$.  On the lattice with
our  parameters and quark mass, it is found that the lightest vector and
scalar meson masses  are $am_v=0.815(5)$ and $am_s=1.39(5)$ respectively
-- see  Ref.~\cite{shan,McN+M}. The fourier transform of these pole
terms is then  a Yukawa form $\propto \exp(- m r)/r$, which -- in
principle -- can now be compared directly with the charge and matter
radial densities measured here. However, it is only the asymptotic form
that should be used in this comparison since that will be controlled by
the lightest particle  mass. But from  Fig.~\ref{Fig6cm.}  it is seen
that the present data only extends up to $r/a=5(4)$ for the
charge(matter) density, corresponding to $mr = 0.4(0.6)$ respectively
which is not large. Furthermore, the charge density is already well
described by a simple exponential up to $r/a=4$ and so  a comparably
good fit over this range by a Yukawa form is ruled out.  As a
compromise, since the charge density data is somewhat better than that
of the matter density, the charge densities at $r/a=3$ and 4 are used to
extract estimates of  $m_v$ and the matter densities at $r/a=2$ and 3
for an estimate of $m_s$. Case 3 leads to $am_v$=0.7(1) and
$am_s=1.4(4)$ -- the corresponding numbers for Case 3' being 0.8(1) and
1.1(1). Even though these estimates are qualitatively consistent with
the lattice results  of Ref.~\cite{shan,McN+M}, it should be added that
they are dependent on the range of $r$-values used.

 This paper should be considered only as the first step in measuring
densities. Many  extensions and refinements are possible:
 \begin{itemize}
\begin{enumerate}
 \item In the $Q\bar{q}$-system, the measurement of the  $P_{1/2}, \
P_{3/2}, \ D_{3/2}, \   D_{5/2},...$ densities corresponding to the
energies extracted in Ref.~\cite{MP98}. For a given orbital angular
momentum, do these correlations show the  degeneracy predicted in
Ref.\cite{Page}?
 \item Off-axis radial correlations. These would check not only
rotational invariance but also the radial correlation for $r/a=5$, which
could then be measured at, for example, $x/a=3, \ y/a=4$ etc. This point is
of particular interest, since in Fig.~\ref{Fig6cm.} there is a hint that
the correlation is lower than that obtained from a simple extrapolation.
Such a lowering is expected, when the linear confining potential begins
to play a role.
 \item The measurement of correlations in the baryonic and
$(Q^2\bar{q}^2)$ system. Are these similar to those in the $(Q\bar{q})$
case -- as is the case  when comparing correlations in few-nucleon
systems?
 \item The replacement of the quenched by unquenched gauge
configurations. This is not expected to have a significant effect on the
charge sum-rule and correlations. However, as discussed in
Refs.~\cite{foster},\cite{Fosterthesis}, for the matter probe 
disconnected contributions arise that are dependent on quenched versus
unquenched. The difference between the two could highlight the effect of
the quark  condensate.
 \item The use of larger $\beta$ values and lattices to check the
 continuum limit  of the present results.
 \item The use of several light quark masses to enable an attempt at 
extrapolating to the realistic non-strange light quark masses.
 \item Other one body operators. In this work only the
charge$(\gamma_4)$ and  matter $(1)$ probes have been studied. However,
others are of interest -- see, for example, Ref.~\cite{Divitiis} where
the pseudo-vector operator $(\gamma _{\mu}\gamma_5)$ is needed for the
$B^*B\pi$ coupling.  
 \end{enumerate}
\end{itemize}

\section{Acknowledgements}
The authors acknowledge useful correspondence and discussions with 
David Richards and Gunnar Bali.
They also wish to thank the Center for Scientific Computing in Espoo,
Finland for making available resources without which this project could
not have been carried out.
 
\appendix
\section{The correlation functions $C(2)$ and $C(3,r)$}
\label{Aformalism}
This appendix gives details of the derivations of Eqs.~\ref{C_2} and
\ref{C_3tot} for the two- and three-point correlation functions $C(2)$
and $C(3)$.
\subsection{Two-point correlation functions $C(2)$}
 Given the  light-quark propagators in Eqs.~\ref{Gauss2} and
\ref{Gauss3} , then the two-point correlation functions $C(2)$ 
 can be expressed in the four ways shown in Fig.~\ref{fig2e}. 
These are  the same upto statistical
errors, but their combination improves the overall measurement.

Case $a$: The $Q\bar{q}$ meson with the propagator in Eq.~(\ref{Gauss2}).

Case $b$: The $Q\bar{q}$ meson with the propagator in Eq.~(\ref{Gauss3}).

Case $c$: The $\bar{Q}q$ meson with the propagator in Eq.~(\ref{Gauss2}).

Case $d$: The $\bar{Q}q$ meson with the propagator in Eq.~(\ref{Gauss3}).

In cases $a$ and $b$, the heavy(static)-quark propagator from site
$({\bf x}, \ t)$ to site $({\bf x'}, \ t+T)$ is simply
\begin{equation}
\label{AGQ}
G_Q({\bf x}, \ t \ ; \ {\bf x'}, \ t+T)=
\frac{1}{2}(1+\gamma_4) U^Q({\bf x},t,T)\delta_{{\bf x},{\bf x'}} ,
 \end{equation}
 where $U^Q({\bf x},t,T)=\prod^{T-1}_{i=0} U_4({\bf x},t+i)$ is the
gauge link product in the time direction. On the other hand, for cases
$c$ and $d$, the heavy(static)-antiquark propagator from site $({\bf x},
\ t)$ to site $({\bf x'}, \ t+T)$ is simply
 \begin{equation}
\label{AGAQ}
 G_{\bar{Q}}({\bf x}, \ t \ ;{\bf x'}, \ t+T)= \frac{1}{2}(1-\gamma_4)
U^{Q\dagger}({\bf x},t,T)\delta_{{\bf x},{\bf x'}}.
 \end{equation}
The general form of a two-point correlation is constructed from a 
heavy-quark propagating from site 
$({\bf x_1}, \ t)$ to site $({\bf x'_1}, \ t+T)$ and a light-quark 
propagating from site 
$({\bf x_2}, \ t+T)$ to site $({\bf x'_2}, \ t)$ i.e. 
\begin{equation}
\label{ATPC}
 C(2,T)=Tr \langle \Gamma^{\dagger} \  G_Q({\bf x}, \ t \ ;{\bf x'}, \
t+T) \Gamma \ G_q({\bf x}, \ t+T \ ;{\bf x'}, \ t) \rangle, 
 \end{equation}
where $\Gamma$ is the spin structure of the heavy-quark light-quark
vertices at $t$ and $t+T$. In this case, since the $B$-meson is a
pseudoscalar, we have $\Gamma = \gamma_5$.
These definitions lead to the four two-point correlation functions
\begin{equation}
\label{AC_2a}
C(2,T,a)=\sum_{d=3,4}\langle U^Q({\bf x},t,T)\psi^*_d({\bf x},t+T)
\phi_d({\bf x},t)\rangle 
\end{equation}
\begin{equation}
\label{AC_2b}
C(2,T,b)=\sum_{d=1,2}\langle U^Q({\bf x},t,T)\phi^*_d({\bf x},t+T)
\psi_d({\bf x},t)\rangle
\end{equation}
\begin{equation}
\label{AC_2c}
C(2,T,c)=\sum_{d=1,2}\langle U^{*Q}({\bf x},t,T)\psi^*_d({\bf x},t)
\phi_d({\bf x},t+T)\rangle 
\end{equation}
\begin{equation}
\label{AC_2d} 
C(2,T,d)=\sum_{d=3,4}\langle U^{*Q}({\bf x},t,T)\phi^*_d({\bf x},t)
\psi_d({\bf x},t+T)\rangle, 
\end{equation}
where  $d$ is the Dirac spin index. We see that $C(2,T,c)^*=C(2,T,b)$
and $C(2,T,d)^*=C(2,T,a)$, so that the sum leads to
\begin{equation}
\label{AC_2}
 C(2,T)=2 \langle   Re\left[U^Q \ [\sum_{e=3,4} \psi^*_e({\bf
x},t+T)\phi_e({\bf x},t)+ \sum_{d=1,2}\phi^*_d({\bf x},t+T)\psi_d({\bf
x},t)   ]\right] \Big\rangle .
 \end{equation}

\subsection{Three-point correlation functions $C(3)$}
\begin{figure}[ht]
\includegraphics{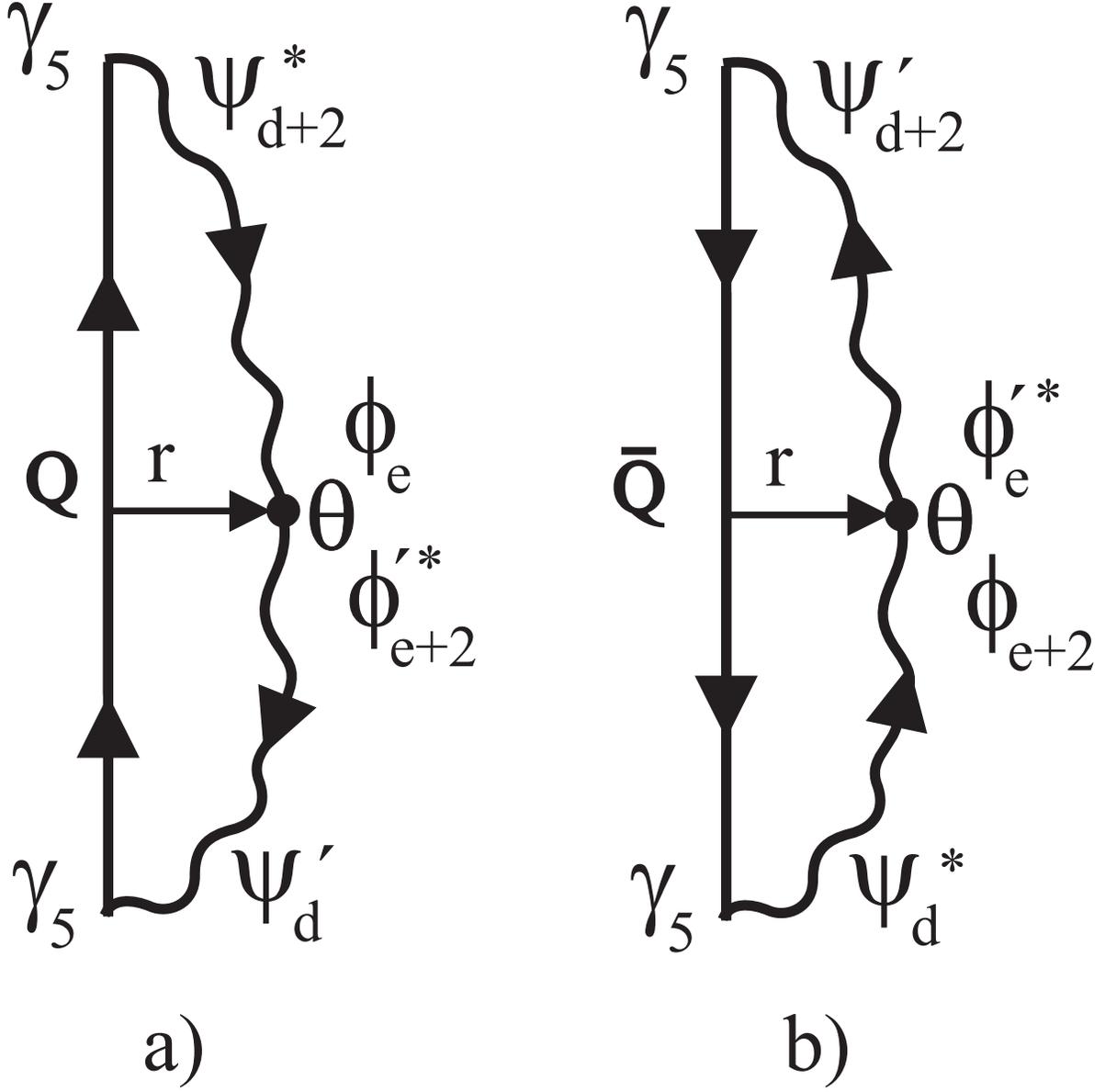}
\caption{The two contributions to the three-point correlation function 
$C(3)$}
\label{fig2r} 
\end{figure}

\begin{figure}[ht]
\includegraphics{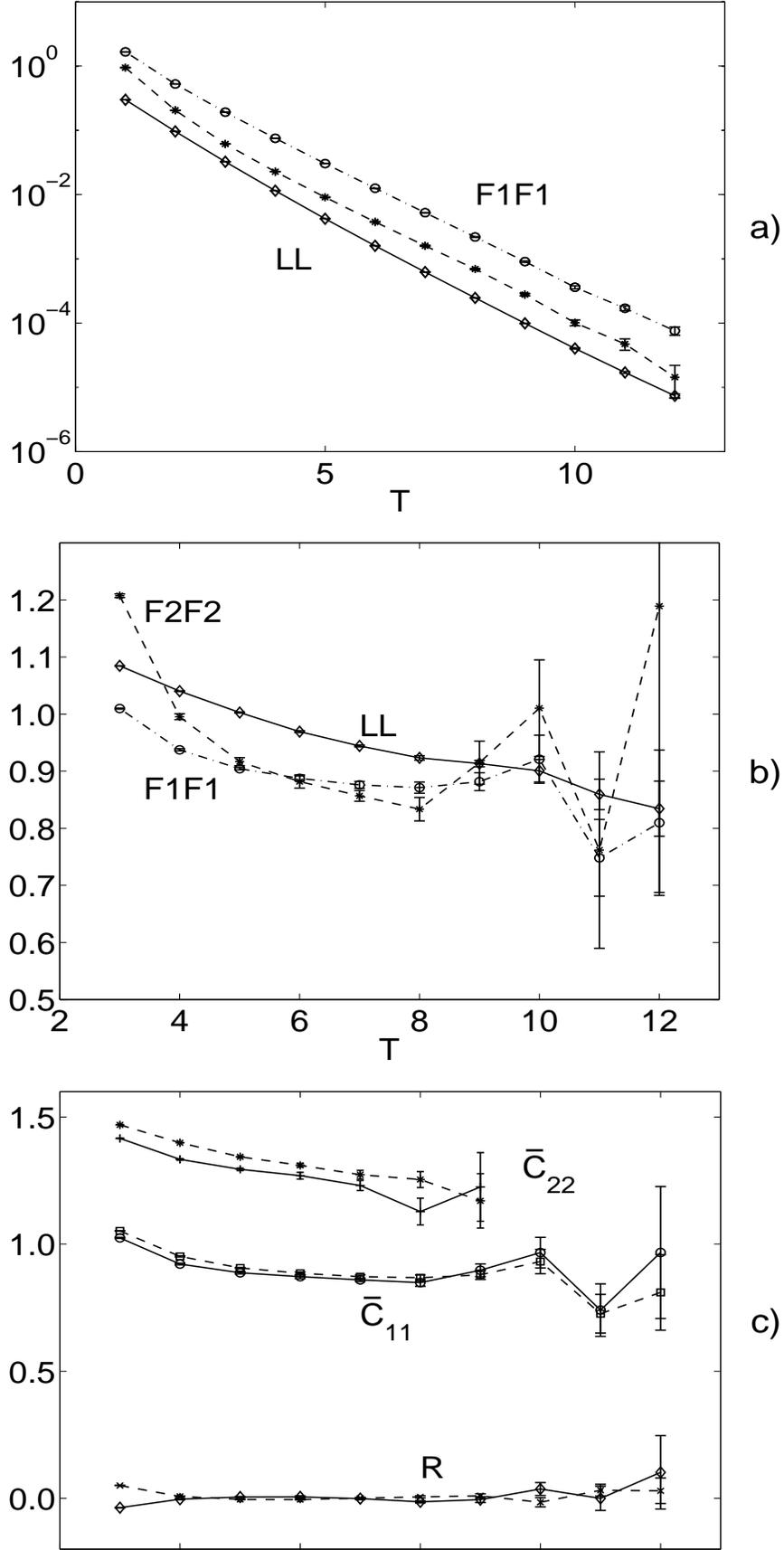}
\caption{a) The basic LL, F1F1 and F2F2 elements of $C(2)$:
b) The values of $E(T)$ for LL, F1F1 and F2F2 separately:
c)  The combinations of L, F1, F2 for cases 3(solid) and 4(dashed) to give
$E[\bar{C}_{\alpha \alpha}]$.
Also shown is the ratio 
$R=\langle \bar{C}_{12}(2,T)/\bar{C}_{11}(2,T) \ \rangle$
defined in Eq.~\protect\ref{Re2}} 
\label{fig3abc}
\end{figure}

\begin{figure}[ht]
\includegraphics{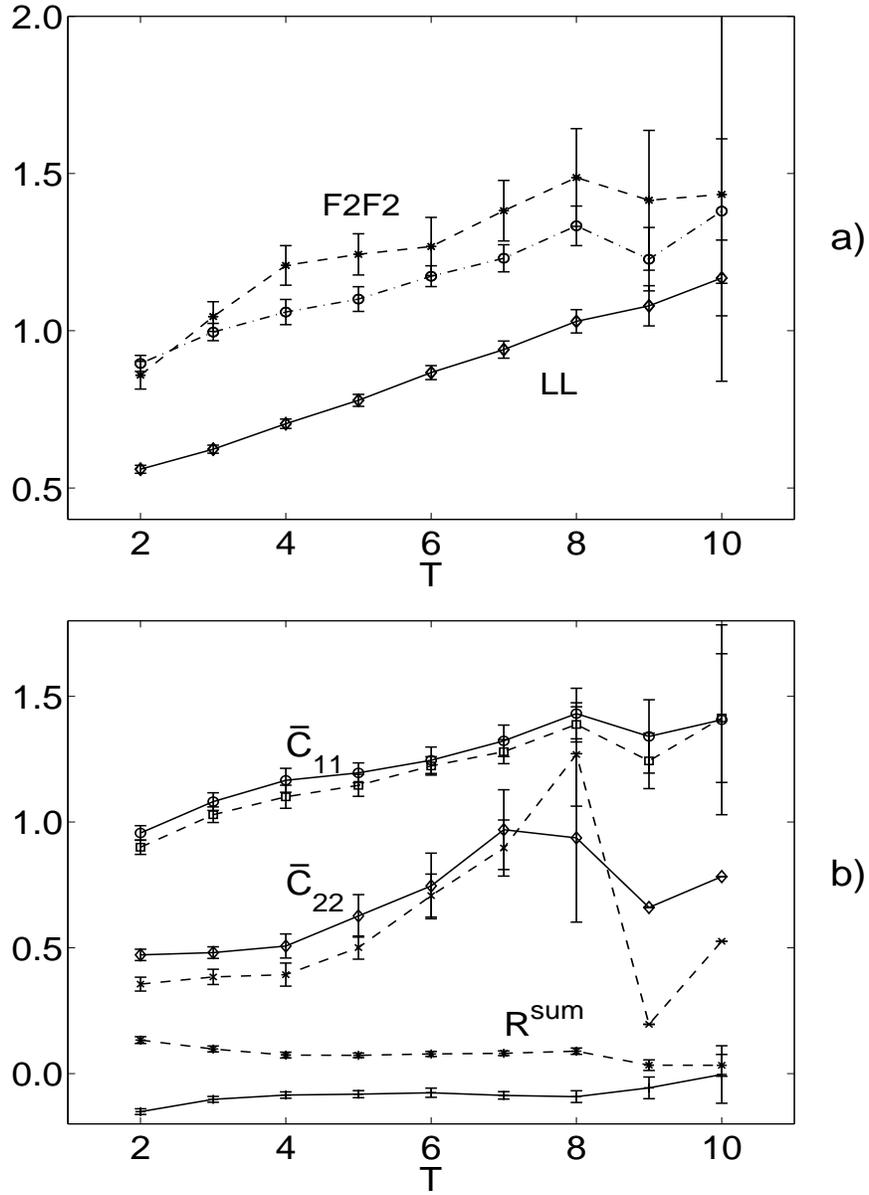}
\caption{ The $\gamma_4$ sum rule: a) Contributions $LL$, $F1F1$ and $F2F2$
separately: b) The combinations of L, F1, F2 for cases 3(solid) and 
4(dashed) to give
$F^{{\rm sum}}[\bar{C}_{\alpha \alpha}(\gamma_4, T)]$ defined in 
Eq.~\protect\ref{sumr} and $R^{{\rm sum}}$
in Eq.~\protect\ref{sumr0}.  }
\label{Fig.4} 
\end{figure}

\begin{figure}[ht]
\includegraphics{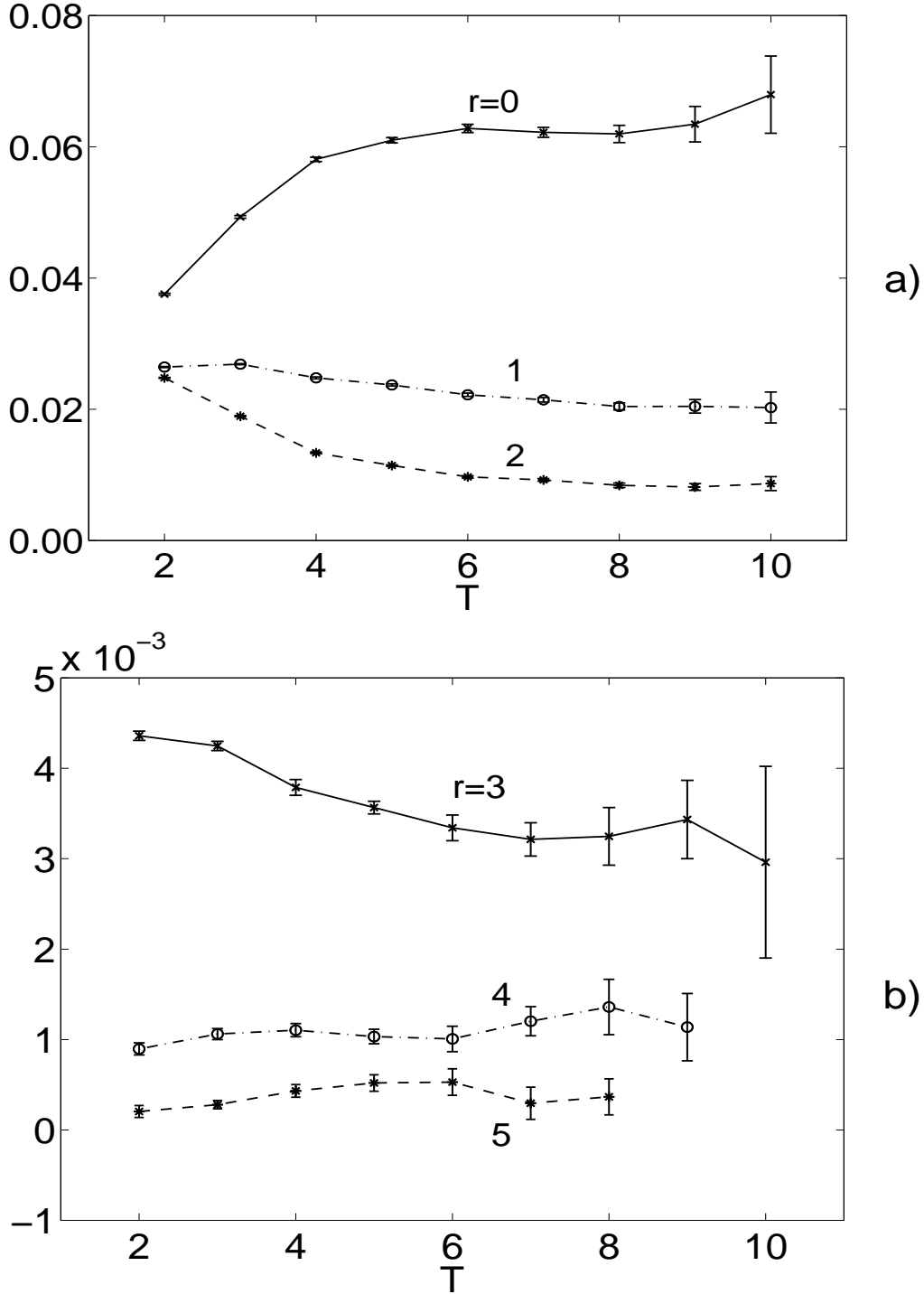}
\caption{ The ratio $\langle C(3,r)/C(2)\rangle$ for Case 3.
a) $r=$ 0,1,2  and b) $r=$ 4,5,6  in lattice units $a\approx 0.17$fm. }
\label{Fig.5} 
\end{figure}

\begin{figure}[ht]
\includegraphics{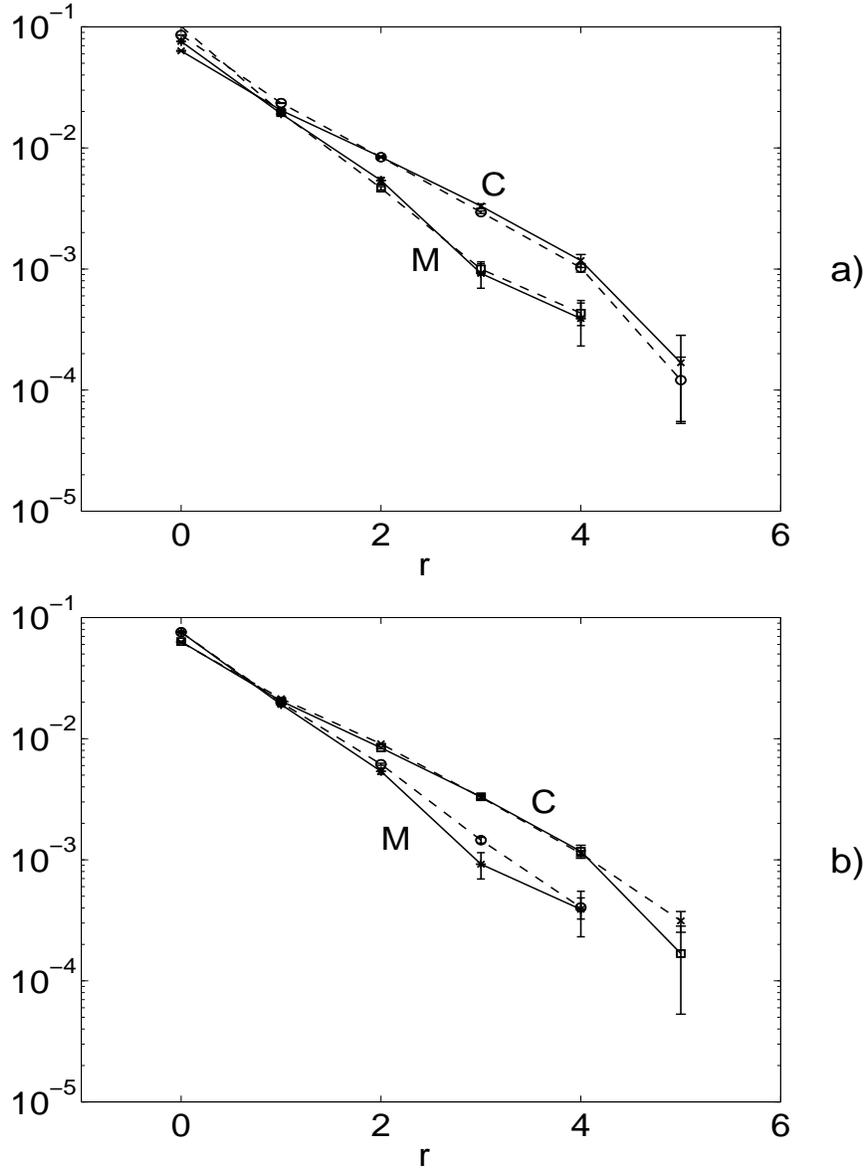}
\caption{ The radial distribution of the ground state charge (C)
and matter(M) densities $x^{11}(r)$. 
a) Case 3(solid) and $3'$(dashed) with $T_{min}(3)=8$:
b) Case 3 for $T_{min}(3)$=8(solid) and 6(dashed).
 The interquark distance $(r)$ is in lattice units $a\approx 0.17$fm.}
\label{Fig6cm.} 
\end{figure}

\begin{figure}[ht]
\includegraphics{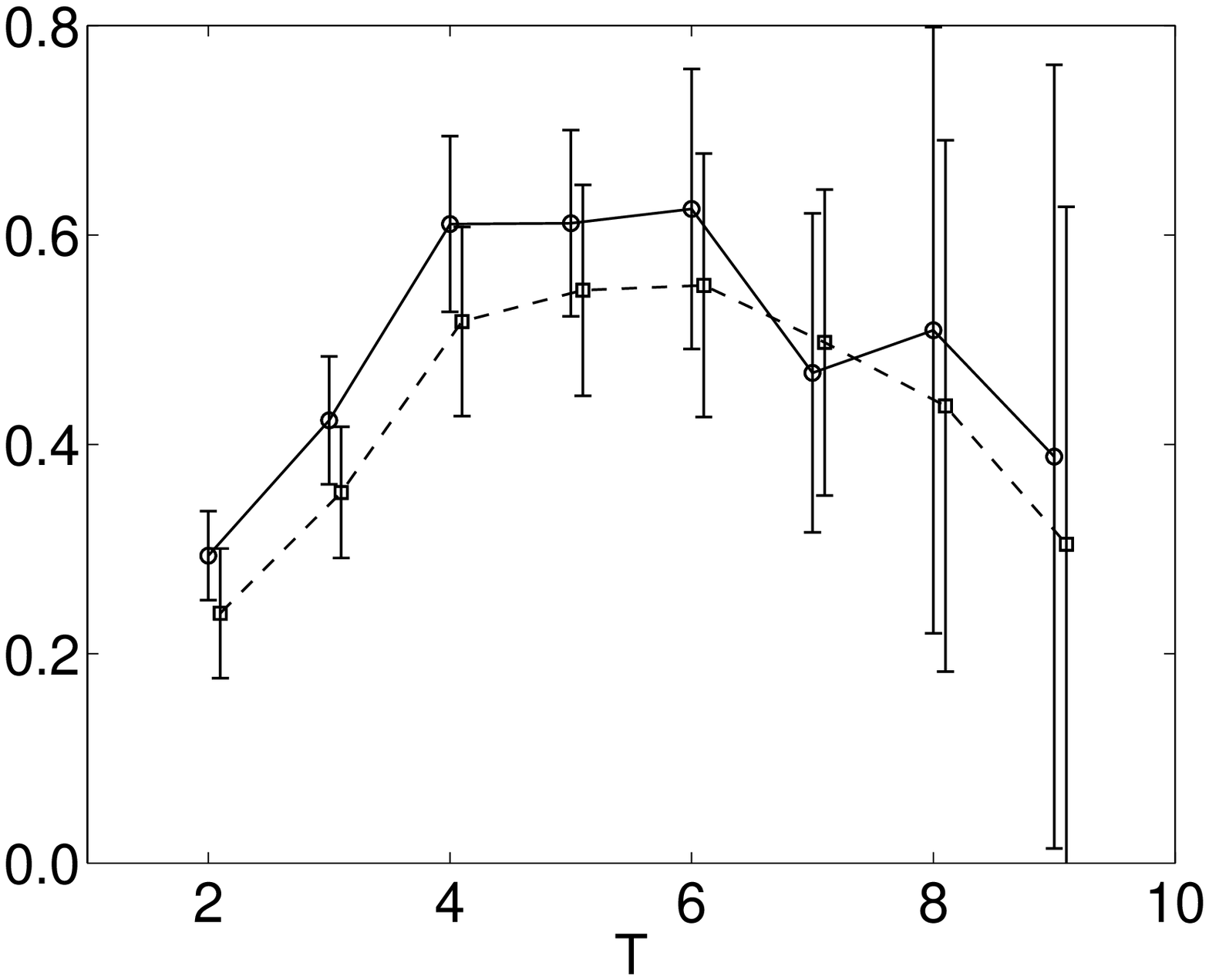}
\caption{ Matter Sum rule with Cases 3(solid) and 4(dashed) 
L,F1,F2 combinations }
\label{Fostersum}
\end{figure}

The construction of the three-point correlation functions $C(3)$
follows that of $C(2)$ and are depicted in Fig.~\ref{fig2r}. 
  
Consider the probe is at $t=0$ and  that $Q$ propagates from
$({\bf x},-t_2)$ to $({\bf x},t_1)$. Analogous to Eq.~(\ref{ATPC}) the 
general form of $C(3)$ --  when involving a $Q$ -- is then   
\begin{equation}
\label{AC_3Q}
 C^Q(3,-t_2, \ t_1, \ {\bf r})=Tr \langle \Gamma^{\dagger} \ G_Q({\bf
x}, \ -t_2 \ ;{\bf x}, \ t_1) \Gamma \ G_q({\bf x}, \ t_1; {\bf x+r},\
0) \ \Theta ({\bf r}) \   G'_q({\bf x+r},\ 0; {\bf x}, \ -t_2) \rangle.
 \end{equation}
This combination of the $G_q$ and $G'_q$ defined in 
Eqs.~(\ref{Gauss2},\ref{Gauss3}) ensures that only the local field
$\phi$ occurs at the probe position ${\bf r}$. When 
$\Theta ({\bf r})=\gamma_4$, this expression reduces to
\[C^Q(3, -t_2, \ t_1, \ {\bf r})=\Big\langle 
\left[\sum_{d=1,2} \ U^Q \psi_{d+2}^*({\bf x}, \ t_1)\psi'_d({\bf x}, \
-t_2) \right] \ \]
\begin{equation}
\label{AC_3Q1}
 \left[\sum_{e=1,2} \ \phi_e({\bf x+r}, \ 0)\phi^{'*}_{e+2}({\bf x+r}, \
0) -\phi_{e+2}({\bf x+r}, \ 0)\phi^{'*}_e({\bf x+r}, \ 0)\right]
\Big\rangle.
 \end{equation}
Care must be exercised here, since $\phi_e({\bf x+r}, \ 0)$ arises from 
$G_q$, whereas $\phi^{'*}_e({\bf x+r}, \ 0)$
is from $G_q'$ and so the two cannot be combined.
 The last equation can now be written as
\begin{equation}
\label{AQXY}
 C^Q(3, -t_2, \ t_1, \ {\bf r})=\langle \ [X^q_R+iX^q_I]\ [Y^Q_R+iY^Q_I]
\ \rangle,
 \end{equation}
where -- supressing color indices --
\[X^q_R=\sum_{e=1,2} \
[\phi_e(R)\phi^{'}_{e+2}(R)+\phi_e(I)\phi^{'}_{e+2}(I)
\mp \phi_{e+2}(R)\phi^{'}_{e}(R) \mp \phi_{e+2}(I)\phi^{'}_{e}(I)] \]
\begin{equation}
\label{AX}
X^q_I=\sum_{e=1,2} \
[\phi_e(I)\phi^{'}_{e+2}(R)-\phi_e(R)\phi^{'}_{e+2}(I)
 \pm \phi_{e+2}(R)\phi^{'}_{e}(I) \mp \phi_{e+2}(I)\phi^{'}_{e}(R)],
\end{equation}
where the upper signs are for  $\Theta ({\bf r})=\gamma_4$ and the lower
ones for $\Theta ({\bf r})=1$. Supressing color and spin indices --
\[Y^Q_R=\psi'(R)U(R)\psi(R)+\psi'(I)U(R)\psi(I)-\psi'(I)U(I)\psi(R)
+\psi'(R)U(I)\psi(I)\]
\begin{equation}
\label{AY}
Y^Q_I=-\psi'(R)U(R)\psi(I)+\psi'(I)U(R)\psi(R)+\psi'(I)U(I)\psi(I)
+\psi'(R)U(I)\psi(R).
\end{equation}

When involving a $\bar{Q}$, the corresponding three-point correlation
function is
\begin{equation}
 \label{AC_3Qb}
 C_3^{\bar{Q}}(-t_2, \ t_1, \ {\bf r})=Tr \langle \Gamma 
G_{\bar{Q}}({\bf x}, \ t_1 \ ;{\bf x}, \ -t_2) \Gamma^{\dagger}G_q({\bf
x}, \ -t_2 \ ; \  {\bf x+r},\ 0) \ \Theta ({\bf r}) \ G'_q({\bf x+r},\
0, {\bf x} \ ;  \ t_1) \rangle,
 \end{equation}
which can be written in the notation of Eq.~(\ref{AQXY}) as
\begin{equation}
\label{AQbXY}
 C^{\bar{Q}}(3, -t_2, \ t_1, \ {\bf r})=\langle  \ [-X^q_R+iX^q_I] \
[Y^Q_R-iY^Q_I] \ \rangle.
 \end{equation}
The overall three-point correlation function is then
\begin{equation}
\label{AC_3tot}
 C(3,-t_2, \ t_1, \ {\bf r})=C^Q(3)\mp C^{\bar{Q}}(3)=2\langle  \
X^q_RY^Q_R - X^q_IY^Q_I\ \rangle,
 \end{equation}
 where the relative sign enters for $\Theta ({\bf r})=\gamma_4$, since
the $q$ and $\bar{q}$ have opposite charges.

\newpage

\begin{table}
\caption{Values of the parameters $am_{\alpha}$ and $ v_i^{\alpha}$,
where $a\approx 0.17$fm is the lattice spacing.
 Cases 1 -- 4 fit the two point correlations [$C(2)$]. }
\begin{tabular}{ccccc}
$am_{\alpha}$&Case 1&Case 2&Case 3& Case 4\\
$ v_i^{\alpha} \ u_i^{\alpha}$&3*3 $T_{min}=3$&2*2 $T_{min}=3$&3*3
$T_{min}=4$&2*2 $T_{min}=4$ \\ \hline
$am_{1}$&0.8849(10)&0.9005(14)& 0.8721(19)&0.8833(27)\\
$am_{2}$&1.2953(63)&1.355(10)&1.263(13)&1.307(20)\\
$am_{3}$&1.99(10)&--&1.94(30)&--\\
$v^1_{L}$& 0.5164(30)&0.5574(41)&0.4847(56)& 0.5149(82)\\
$v^1_{F1}$&1.5892(48)&1.6761(52)&1.519(10)&1.589(13)\\
$v^1_{F2}$&0.8651(22)&--&0.8402(38)&-- \\
$v^2_{L}$&0.8123(61)&--0.8226(74)&0.816(16)& --0.834(19)\\
$v^2_{F1}$&0.435(22)&--0.065(29)&0.644(49)&--0.373(57)\\
$v^2_{F2}$&--0.393(18)&--& --0.251(33)&--\\
$v^3_{L}$& --0.258(63)&--&--0.28(22)&--\\
$v^3_{F1}$&1.93(32)&--& 2.2(1.4)&--\\
$v^3_{F2}$& --1.40(24)&--&--1.13(81)&--\\ \hline
$n_{data}(2)$&60&30&54&27\\
$n_{param}(2)$&12&6&12&6\\
$n_{dof}(2)$&48&24&42&21\\
$\chi^2/n_{dof}(2)$&3.1&7.5&0.65&1.15\\ \hline
$u^1_{L}$&0.0421&--0.0480&--0.0608&--0.3294\\
$u^1_{F1}$&0.3546&0.6126&0.3258&0.7361\\
$u^1_{F2}$&0.4793&--&0.6364&--\\
$u^2_{L}$&1.1135&--1.2482&1.1575&--1.4024\\
$u^2_{F1}$&--0.1432&0.4151& --0.1025&0.4545\\
$u^2_{F2}$&--0.4016&--&--0.4825&--\\
\end{tabular}
\label{Tablecs}
\end{table}

\newpage
\begin{table}
\caption{Values of the $x^{\alpha \beta}$ for Cases 3 and $3'$.
In Case 3 the 12 parameters describing $C(2)$ are fixed and only the 6 
$x^{\alpha \beta}$ are varied to fit $C(3)$. In Case $3'$ only the 3 
$x^{\alpha \alpha}$ are varied.  
The superfix $^*$ implies
that the number fixed at this value}
\begin{tabular}{lcccccc}
&&Case 3&&&Case $3'$&\\  \hline
$\chi^2(2)/n_{dof}(2)$&&0.65$^*$&&&0.65$^*$&\\
$T_{min}(3)$&4&6&8&4&6&8\\
$\chi^2(3)/n_{dof}(3)$&1.20&0.72&0.26&4.2&2.4&0.68\\ \hline
$x^{11}$&1.26(2)&1.33(2)&1.41(5)&1.12(1)&1.19(1)&1.29(3)\\
$x^{12}$&--0.32(3)&--0.44(7)&--0.5(2)&0$^*$&0$^*$&0$^*$\\
$x^{13}$&0.03(9)&0.0(3)&0(1)&0$^*$&0$^*$&0$^*$\\
$x^{22}$&0.65(7)&0.9(2)&0.9(9)&0.00(4)&--0.3(1)&--1.2(4)\\
$x^{23}$&0.2(2)&0(1)&0(8)&0$^*$&0$^*$&0$^*$\\
$x^{33}$&0.2(9)&4(8)&--&--0.3(8)&1(7)&--\\ \hline
&&Case 4 &&&Case $4'$&\\  \hline
$\chi^2(2)/n_{dof}(2)$&&1.15$^*$&&&1.15$^*$&\\
$T_{min}(3)$&4&6&8&4&6&8\\
$\chi^2(3)/n_{dof}(3)$&1.23&0.78&0.21&3.4&2.0&0.58\\ \hline
$x^{11}$&1.24(2)&1.29(3)&1.36(7)&1.10(1)&1.17(2)&1.26(3)\\
$x^{12}$&--0.36(5)&--0.45(11)&--0.5(3)&0$^*$&0$^*$&0$^*$\\
$x^{22}$&0.65(13)&0.8(4)&1(2)&--0.19(5)&--0.6(2)&--1.6(6)\\
\end{tabular}   
\label{Tablexs}
\end{table}

\newpage
\begin{table}
\caption{The ground and first excited state charge densities $x^{11}$,
$x^{22}$ for different values of $T_{min}(3)$.
In Case 3 only the  six $x^{\alpha \beta}(r)$ are varied
-- the ${\bf v}$-vectors being those determined earlier from fitting
$C(2)$ and shown in Table~\protect\ref{Tablecs}. In Case $3'$
only the three $x^{\alpha \alpha}(r)$ are varied. The numbers in [...]
are the $\chi^2(3)/n_{dof}(3)$. The entries denoted by '--' have
unreasonably large $\chi^2(3)/n_{dof}(3)$.  The interquark distance
$r$ is in lattice units  of $a\approx 0.17$fm.}
\begin{tabular}{c|cccccc}
$x^{11}$,Case 3&$r=0$&$r=1$&$r=2$&$r=3$&$r=4$&$r=5$\\

$T_{min}(3)=6$&0.0627(2) &0.0214(1) &0.00906(8)&0.00328(6)&0.00112(6)
              &0.00031(6)\\
              &[0.37]    & [2.54]   & [2.37]   & [0.20]   & [0.40]
              & [0.82]   \\
$T_{min}(3)=7$&0.0629(3) &0.0209(2) &0.00882(10)&0.00326(9)&0.00120(9)
              &0.00021(8)\\
              &[0.27]    & [1.67]   & [1.49]   & [0.25]   & [0.34]
              & [0.59]   \\
$T_{min}(3)=8$&0.0633(5)  &0.0204(2)  &0.0084(2)&0.0033(1)&0.0012(1)
              &0.00017(11)\\
              &[0.16]   & [1.38]  & [0.36]   & [0.33]   & [0.49]
              &[0.68]   \\
$T_{min}(3)=9$&0.0631(7)  &0.0200(4)  &0.0081(2)&0.0033(2)&0.0010(2)
              &--0.0002(2)\\
              &[0.24]   & [1.48]  & [0.05]   & [0.48]   & [0.53]
              &[0.39]      \\ \hline

$x^{11}$, Case $3'$&$r=0$&$r=1$&$r=2$&$r=3$&$r=4$&$r=5$\\
$T_{min}(3)=7$&--&-- &0.00856(5)&0.00287(4)&0.00100(5)
              &0.00018(5)\\
              &--    & --   & [2.09]   & [1.87]   & [0.94]
              & [0.58]   \\
$T_{min}(3)=8$&--   &--  &0.0084(1)&0.0029(1)&0.0010(1)
              &0.00012(7)\\
              &--   & -- & [0.38]   & [1.08]   & [0.67]
              &[0.81]   \\
$T_{min}(3)=9$&--   &--  &0.0082(1)&0.0029(1)&0.0009(1)
              &0.0000(1)\\
              &--   & -- & [0.07]   & [0.80]   & [0.37]
              &[0.33]   \\ \hline \hline
$x^{22}$, Case 3&$r=0$&$r=1$&$r=2$&$r=3$&$r=4$&$r=5$\\
$T_{min}(3)=6$&0.192(3)&0.011(1)&0.0019(7)&0.0009(5)&0.0004(6)&0.0003(6)\\
$T_{min}(3)=8$&0.19(1) &0.010(4)&0.001(3) &0.001(2) &0.000(3) &0.001(2)\\
$x^{22}$, Case 3$'$&$r=0$&$r=1$&$r=2$&$r=3$&$r=4$&$r=5$\\
$T_{min}(3)=7$&--&--&--0.001(1)&--0.0033(4)&--0.0020(6)&0.000(1)\\ 
\end{tabular}
\label{Table.xr33p}
\end{table}

\newpage  

\begin{table}   
\caption{The parametrization of the wavefunction as
$w_c(r)=A_c\exp[-(r/r_c)^p]$ for various values of $T_{min}(3)$
and  ranges of $r$. Also given is $I=\pi A_c^2r_c^3$.
Here only fits with $\chi^2/dof \approx 1$.
The cases with $p=1^*$ have $p$ fixed. Both
$A_c$ and $r_c$ are expressed in terms of lattice units $a\approx 0.17$fm. }
\begin{tabular}{cccccc}
Case 3&$r-$range&$A_c$&$r_c$ &$p$& $I$ \\ \hline
$T_{min}(3)=7$& 2--4&0.254(7)&2.01(5)&1$^*$&1.52(14)\\
              & 1--4&0.199(5)&2.87(14)&1.23(5)&1.52\\
$T_{min}(3)=8$& 1--4&0.225(3)&2.20(4)&1$^*$&1.71(10)\\
              & 2--4&0.238(10)&2.10(8)&1$^*$&1.65(23)\\    
              & 1--4&0.208(11)&2.59(25)&1.13(9)&1.56\\  
$T_{min}(3)=9$& 1--4&0.224(4)&2.18(6)&1$^*$&1.64(15)\\
              & 1--4&0.202(15)&2.66(37)&1.16(14)&1.47\\  
              & 1--5&0.196(12)&2.81(31)&1.22(12)&1.41\\ \hline  
Case $3'$&&&&\\ \hline
$T_{min}(3)=7$& 2--4&0.275(4)&1.84(2)&1$^*$&1.47(6)\\
              & 1--4&0.249(5)&2.12(8)&1.08(3)&1.43\\
$T_{min}(3)=8$& 1--4&0.257(2)&1.94(2)&1$^*$&1.50(5)\\
              & 2--4&0.261(6)&1.91(4)&1$^*$&1.49(12)       \\
              & 1--4&0.250(9)&2.04(13)&1.03(4)&1.47 \\

$T_{min}(3)=9$& 1--4&0.251(2)&1.96(3)&1$^*$&1.49(7)\\
              & 2--4&0.260(10)&1.90(7)&1$^*$&1.46(20)\\
              & 1--4&0.237(12)&2.19(21)&1.08(7)&1.41\\

\end{tabular}
\label{Table.Ar0a}
\end{table}

\vskip 1cm

\begin{table}
\caption{Values of the $x^{\alpha \beta}$ in Cases 3 and $3'$ for the 
{\em matter} distribution. Notation as in Table \protect\ref{Tablexs}.}
\begin{tabular}{lcccccc}
&&Case 3&&&Case $3'$&\\  \hline
$\chi^2(2)/n_{dof}(2)$&&0.65$^*$&&&0.65$^*$&\\
$T_{min}(3)$&4&6&8&4&6&8\\
$\chi^2(3)/n_{dof}(3)$&0.40&0.45&0.35&0.72&0.43&0.29\\ \hline
$x^{11}$&0.56(3)&0.49(6)&0.38(15)&0.48(2)&0.45(3)&0.35(7)\\
$x^{12}$&--0.22(7)&--0.1(2)&--0.1(5)&0$^*$&0$^*$&0$^*$\\
$x^{13}$&--0.2(2)&0.0(8)&--1(5)&0$^*$&0$^*$&0$^*$\\
$x^{22}$&0.4(2)&0.4(5)&1(2)&--0.11(9)&0.0(3)&0(1)\\
$x^{23}$&0.3(4)&0(3)&--&0$^*$&0$^*$&0$^*$\\
$x^{33}$&2(2)&--&--&2(2)&--&--\\ \hline
\end{tabular}   
\label{Tablefxs}
\end{table}

\newpage
\begin{table}
\caption{ The Matter densities with 
notation as in Table~\protect\ref{Table.xr33p}}
\begin{tabular}{c|cccccc}
$x^{11}$,Case 3&$r=0$&$r=1$&$r=2$&$r=3$&$r=4$&$r=5$\\
$T_{min}(3)=6$&0.0758(3)&0.0199(1)&0.0062(1)&0.00145(8)&0.00040(8)
              &0.00004(7) \\
              & [0.20]  & [1.07]  & [0.83 ] & [0.83]   & [0.23]
              & [0.60]  \\
$T_{min}(3)=8$&0.0758(7)&0.0192(3)&0.0054(3)&0.0009(3)&0.0004(2)
              &0.0005(2) \\
              &[0.25]   &[0.04]   & [0.19]  & [0.94]  &[0.19]
              &[0.06]   \\ \hline
$x^{11}$,Case 3$'$&$r=0$&$r=1$&$r=2$&$r=3$&$r=4$&$r=5$\\
$T_{min}(3)=6$&--&--&--&0.00113(4)&0.00034 (4)
              &0.00003(5) \\
              & --  & --  & -- & [1.82]   & [0.24]
              & [0.57]  \\
$T_{min}(3)=8$&--&0.0196(2)&0.0046(2)&0.0010(1)&0.00043(9)
              &0.0003(1)\\
              &-- &[1.24]   & [0.80]  & [0.77]  &[0.16]
              &[0.16]   \\ \hline
\end{tabular}
\label{Table.matter}
\end{table}

\vskip 1cm


\begin{table}
\caption{The parametrization of the matter distribution wavefunction as \newline
 $w_m(r)=A_m\exp[-r/r_m]$ -- notation as in 
Table~\protect\ref{Table.Ar0a} }
\begin{tabular}{ccccc}
Case 3&$r$ range&$A_m$&$r_m$&$I_m$\\ \hline
$T_{min}(3)=6$&2--4&0.33(2)&1.40(5)& 0.92(14) \\
$T_{min}(3)=8$&0--4&0.275(1)&1.46(2)&0.75(3) \\
&1--4&0.271(8)&1.49(5)&0.77(9) \\
&2--4&0.37(8)&1.24(16)&0.81(46) \\ \hline
Case 3$'$&$r$ range&$A_m$&$r_m$&$I_m$\\ \hline
$T_{min}(3)=6$&1--4&0.293(2)&1.39(1)& 0.72(2) \\
$T_{min}(3)=8$&1--4&0.287(5)&1.39(3)&0.69(5) \\ \hline
\end{tabular}
\label{Table.matterAr0}
\end{table}

\newpage

\newpage

\newpage
\thispagestyle{empty}

\newpage

\newpage

\newpage

\newpage

\begin{thebibliography}{99}
\bibitem{Divitiis} G.M.~de Divitiis, L.~Del Debbio, M.~Di Pierro,
J.M.~Flynn, C.~Michael and J.~Peisa, JHEP 9810:010,1998 {\tt
hep-lat/9807032}
\bibitem{foster}
UKQCD Collaboration, M.~Foster and C.~Michael,
Phys. Rev. {\bf D59}, 074503 (1999), {\tt hep-lat/9810021}
\bibitem{Fosterthesis} M.~Foster,  University of Liverpool PhD thesis 1998
\bibitem{models}T.A.~L\"{a}hde, C.J.~Nyf\"{a}lt and D.O.~Riska, Nucl.
Phys. {\bf A 674}, 141 (2000), {\tt hep-ph/9908485}  
 
M. Di Pierro and E. Eichten,  {\tt hep-ph/0104208}
\bibitem{Gl} W. Gl\"{o}ckle and H. Kamada, 
Phys. Rev. Lett. {\bf 71} 971 (1993) and

M. Baldo et al., NuPECC Report  June 2000

\bibitem{fields} P.~Pennanen, A.~M.~Green and C.~Michael,
Phys.\ Rev.\ D {\bf 59} 014504 (1999), {\tt hep-lat/9804004}; 
Phys.\ Rev.\ D {\bf 56} 3903 (1997), {\tt hep-lat/9705033};
G.~S.~Bali, K.~Schilling and C.~Schlichter,
Phys.\ Rev.\ D {\bf 51}, 5165 (1995), {\tt hep-lat/9409005}.



\bibitem{MP98} C.~Michael and J.~Peisa, Phys.Rev. {\bf D58}, 034506 (1998),
{\tt hep-lat/9802015}.

\bibitem{GMP93} A.M.~Green, C.~Michael and J.E.~Paton, Nucl.Phys.
 {\bf A554}, 701 (1993), {\tt hep-lat/9209019}.


\bibitem{McN+M} 
UKQCD Collaboration, C.~McNeile and C.~Michael, {\tt hep-lat/0010019}

\bibitem{bowler} K.C.~Bowler, L.~Del Debbio, J.M.~Flynn, G.N.~Lacagnina,
V.I.~Lesk, C.M.~Maynard and D.G.~Richards, {\tt hep-lat/0007020}

\bibitem{GLPM}  A.M.~Green, J.~Lukkarinen,P.~ Pennanen and C.~Michael,
Phys.Rev. {\bf D53}, 261 (1996), {\tt hep-lat/9508002}.


\bibitem{MP99} UKQCD Collaboration, C.~Michael and P.~ Pennanen,
Phys.Rev.{\bf D60}, 054012 (1999) {\tt hep-lat/9901007}; 
UKQCD Collaboration, C.~Michael, P.~Pennanen and A.M.~Green, 
Proceedings of the 17th International
Symposium on Lattice Field Theory (LATTICE 99), Pisa, Italy,
Nucl. Phys. B(Proc. Suppl.) {\bf 83-84}, 200 (2000), {\tt
hep-lat/9908032};\\
P. Pennanen and C. Michael, {\tt hep-lat/0001015}.
\bibitem{B+D} J.D.~Bjorken and S.D.~Drell, Relativistic Quantum Mechanics
(McGraw-Hill, Inc., 1964)
\bibitem{shan} UKQCD Collaboration, H. Shanahan et al.,  Phys. Rev. {\bf D55}
(1997) 1548
\bibitem{Page} P.R. Page, T. Goldman and J. N. Ginocchio, {\tt hep-ph/0002094}

\end{thebibliography}
\end{document}